\documentclass[
reprint,
superscriptaddress,
amsmath,amssymb,
aps,
prb,
]{revtex4-2}

\usepackage[dvipsnames]{xcolor}
\usepackage{graphicx}
\usepackage[caption=false]{subfig}
\usepackage{dcolumn}
\usepackage{bm}
\usepackage{hyperref}
\hypersetup{
  colorlinks,
  citecolor=blue,
  linkcolor=blue,
  urlcolor=blue}

\newcommand{\beginsupplement}{
        \setcounter{table}{0}
        \renewcommand{\thetable}{S\arabic{table}}
        \renewcommand{\theHtable}{S\thetable}
        \setcounter{figure}{0}
        \renewcommand{\thefigure}{S\arabic{figure}}
        \renewcommand{\theHfigure}{S\thefigure}
        \setcounter{equation}{0}
        \renewcommand{\theequation}{S\arabic{equation}}
}

\begin{document}

\title{Non-equilibrium dynamics of electron emission from cold and hot graphene under proton irradiation}

\author{Yifan Yao}
\affiliation{Department of Materials Science and Engineering, University of Illinois at Urbana-Champaign, Urbana, IL 61801, USA}

\author{Alina Kononov}
\affiliation{Center for Computing Research, Sandia National Laboratories, Albuquerque, NM 87123, USA}

\author{Arne Metzlaff}
\affiliation{University of Duisburg-Essen, Faculty of Physics and CENIDE, 47057 Duisburg, Germany}

\author{Andreas Wucher}
\affiliation{University of Duisburg-Essen, Faculty of Physics and CENIDE, 47057 Duisburg, Germany}

\author{Lukas Kalkhoff}
\affiliation{University of Duisburg-Essen, Faculty of Physics and CENIDE, 47057 Duisburg, Germany}

\author{Lars Breuer}
\affiliation{University of Duisburg-Essen, Faculty of Physics and CENIDE, 47057 Duisburg, Germany}

\author{Marika Schleberger}
\affiliation{University of Duisburg-Essen, Faculty of Physics and CENIDE, 47057 Duisburg, Germany}

\author{Andr\'e Schleife}
\email{schleife@illinois.edu}
\affiliation{Department of Materials Science and Engineering, University of Illinois, Urbana-Champaign, Urbana, IL 61801, USA}
\affiliation{Materials Research Laboratory, University of Illinois at Urbana-Champaign, Urbana, IL 61801, USA}
\affiliation{National Center for Supercomputing Applications, University of Illinois at Urbana-Champaign, Urbana, IL 61801, USA}

\begin{abstract}
Characteristic properties of secondary electrons emitted from irradiated two-dimensional materials arise from multi-length and time-scale relaxation processes that connect the initial non-equilibrium excited electron distribution with their eventual emission.
To understand these processes, which are critical for using secondary electrons as high-resolution thermalization probes, we combine first-principles real-time electron dynamics with modern experiments.
Our data for cold and hot proton-irradiated graphene shows signatures of kinetic and potential emission and generally good agreement for electron yields between experiment and theory.
The duration of the emission pulse is about 1.5 femtoseconds, indicating high time resolution when used as a probe.
Our newly developed method to predict kinetic energy spectra shows good agreement with electron and ion irradiation experiments and prior models.
We find that lattice temperature significantly increases secondary electron emission, whereas electron temperature has a negligible effect.
\end{abstract}

\maketitle

Secondary electrons can be emitted from the surface of a target material upon impact of ions or electrons.
The spectral and spatial distribution of the emitted electrons play a crucial role in modern light-ion microscopy since they provide high-resolution surface morphology images of target materials with minimal collateral damage~\cite{fox_helium_2013, iberi_maskless_2015, hlawacek_helium_2014}. 
The electronic stopping power and charge exchange between a projectile ion and the target material can be experimentally probed, e.g.\ using ion transmission microscopy.
However, the emerging electron and ion dynamics within the target after irradiation and its impact on the intensity and kinetic energy distribution of emitted secondary electrons remains elusive in experiments. 
Hence, emitted secondary electrons can provide important insight e.g.\ into the electron and ion response of the target material \cite{werner_secondary_2020, hlawacek_helium_2014, niggas_ion-induced_2022}.

Achieving such insight requires a deep understanding of secondary electron emission as a complex multi-length and time-scale process that emerges from the interaction between the incident projectile ion and the target material. 
It includes the dynamics of the projectile charge state, the secondary electron emission probability, and the electron-electron and electron-ion relaxation dynamics following the initial excitation.
An analytical model of the secondary electron kinetic energy spectra by Chung \cite{chung_simple_1974} is based on the work function as a barrier against emission and the electron-electron scattering mean free path as the probability for excited electrons to diffuse from bulk to the surface.
This model has successfully described the kinetic energy distribution of secondary electrons emitted from metals under electron irradiation. 
Later, a modified version was introduced to describe the kinetic energy distribution of secondary electrons in helium ion microscopy \cite{petrov_secondary_2011}. 
While these models rely on the static work function of the target material as a single empirical parameter, for a highly excited surface~\cite{niggas_ion-induced_2022} the surface potential and, thus, the work function is strongly perturbed.

First-principles theoretical efforts \cite{bernardi_ab_2014,brown_nonradiative_2016,brown_experimental_2017} focus on the relaxation dynamics in the target via thermalization of the radiation-induced excited electrons towards a Fermi-Dirac distribution through electron-electron scattering, and the subsequent emergence of thermal equilibrium between nuclei and electrons through electron-phonon scattering.
With the development of modern ion beam experiments~\cite{Breuers_Concept_2019, KUCHARCZYK_Simulation_2020, Golombek_180short_2021, kalkhoff_ultrashort_2023, Mihaila_ultrahsort_2023}, the time interval between successive ion impacts on a sample surface is reaching the few picosecond regime, i.e., prior to electron thermalization.
Ion impact on pre-excited material can influence the secondary electron emission.
Hence, a direct real-time electron dynamical simulation that describes the influence of thermalization on secondary electron emission during successive ion impacts is desirable but still missing.

Non-equilibrium dynamics of secondary electron emission becomes even more complicated for two-dimensional (2D) systems, where the impacting ion does not reach an equilibrium charge state before leaving the target, leading to a response that differs from bulk.
For bulk targets, electronic stopping has been extensively investigated~\cite{ullah_core_2018,ullah_electronic_2015,shukri_ab_2016, lee_hot-electron-mediated_2019,correa_calculating_2018, schleife_accurate_2015} and for self-irradiated silicon, first-principles simulations have shown that a highly charged ion equilibrates its charge state within a few nanometers \cite{lee_multiscale_2020}. 
For single or few-layer materials, energy deposition rates~\cite{kononov_pre-equilibrium_2020}, charge excitation~\cite{kononov_anomalous_2021,kononov_pre-equilibrium_2020}, and charge capture processes~\cite{vazquez_electron_2021,kononov_pre-equilibrium_2020} deviate from bulk behavior.
Theoretical models for the escape probability of excited electrons for bulk, such as Chung's model~\cite{chung_simple_1974}, rely on the electron-electron mean free path.
However, this quantity is highly anisotropic in 2D systems and the electron mean free path transverse to the graphene layer is comparable to the layer thickness~\cite{geelen_nonuniversal_2019}.
Besides, secondary electrons in bulk undergo scattering and can reverse their momentum, which may not occur in thin targets~\cite{kononov_first-principles_2022}, due to the short mean free path.
First-principles simulations can provide insight into such complex electron-ion dynamics and lead to improved understanding of ion irradiation of 2D or thin-film materials.

At the same time, the challenge of grasping such electron dynamics also drives the advancement of new experimental methods.
While ion-surface or ion-solid interactions have been studied for decades, direct experimental investigations into the dynamics induced by ion impact within the target material still remain unfeasible. 
Only the initial state before impact and the final state several nanoseconds after impact are directly measurable.
Pump-probe experiments can bridge this gap for laser excitation experiments \cite{Elzinga_PumpProbe_1987, Sokolowski-Tinten_PumpProbe_2004},
where a single laser beam is split to establish a fixed temporal relationship between the resulting pump and probe pulses. 
Varying the delay interrogates the induced dynamics in the sample and lasers with attosecond pulse durations \cite{lHuillier_Nobel_2006, Hentschel_Nobel_2001, Agostini_Attosecond_2004} enable examining sub-femtosecond processes. 
For ion pulses, their charge and mass constrain the reduction of monoenergetic pulse durations.
However, recent developments allow experiments with few-picosecond ion pulses \cite{Breuers_Concept_2019, KUCHARCZYK_Simulation_2020, Golombek_180short_2021, kalkhoff_ultrashort_2023, Mihaila_ultrahsort_2023}, 
making electronic excitation time scales in 2D materials, such as graphene, experimentally accessible.

Here, we develop a computational first-principles description of the secondary electron emission dynamics, including duration and kinetic energy spectra, for proton-irradiated graphene and make connections with experiments.
We use real-time time-dependent density functional theory (RT-TDDFT) and Ehrenfest dynamics, which have successfully simulated femtosecond electron dynamics under external irradiation\cite{niggas_ion-induced_2022,kononov_first-principles_2022,kononov_pre-equilibrium_2020,vazquez_electron_2021,gruber_ultrafast_2016,ueda_quantum_2016,ueda_secondary-electron_2018,tsubonoya_time-dependent_2014}, providing temporal and spatial resolutions that are difficult to achieve experimentally. 
We explicitly consider the effects of non-vanishing electron and lattice temperatures of pre-excited graphene, e.g.\ caused by preceding ion impact, advancing understanding of secondary electrons as a probe for thermalization processes.
We provide comprehensive characterization of secondary electron pulses from ground-state and excited monolayer graphene, as a prototypical 2D material, and critical insights into its local and temperature-dependent material properties.

\begin{figure}
\includegraphics[width=0.95\columnwidth]{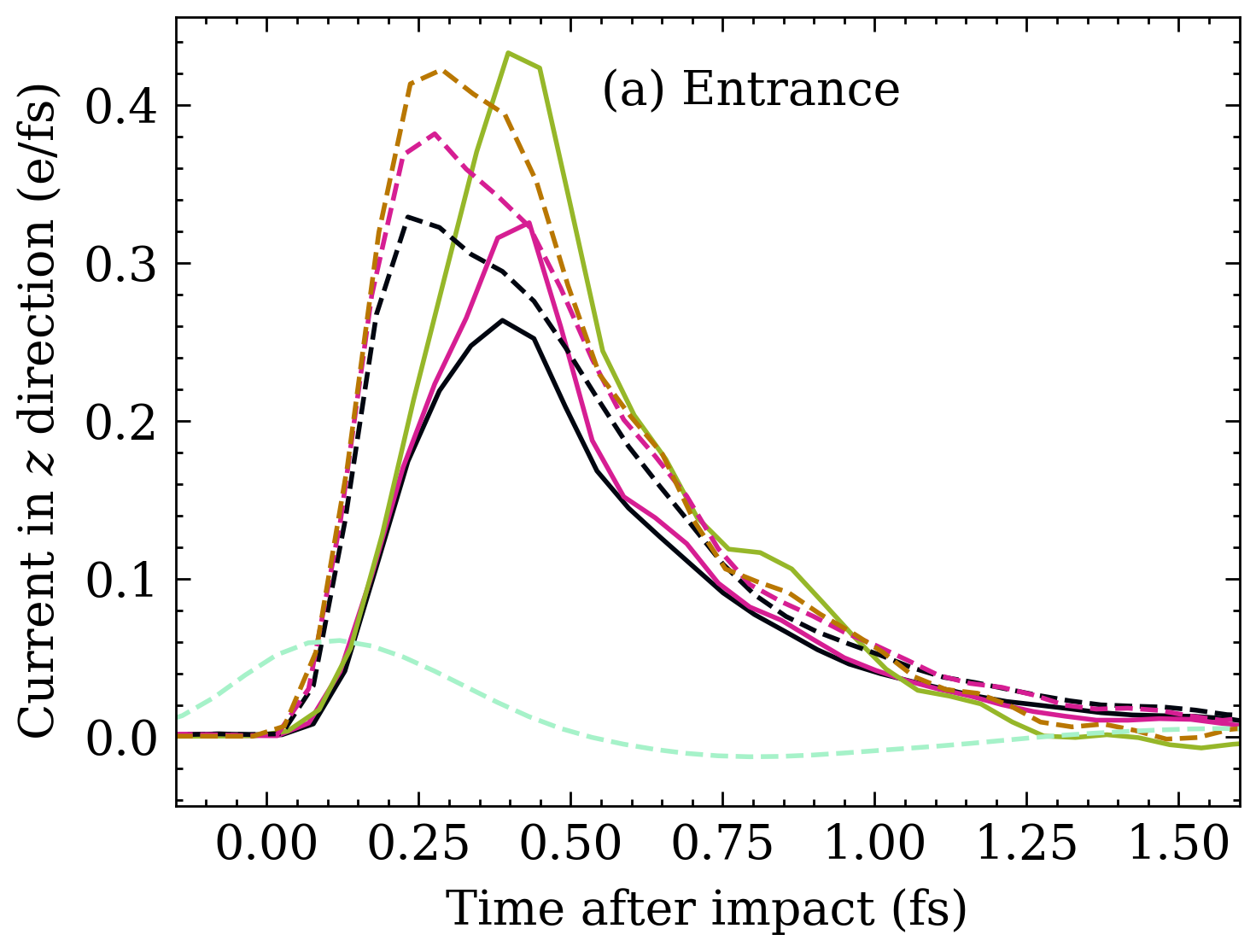}\\
\includegraphics[width=0.95\columnwidth]{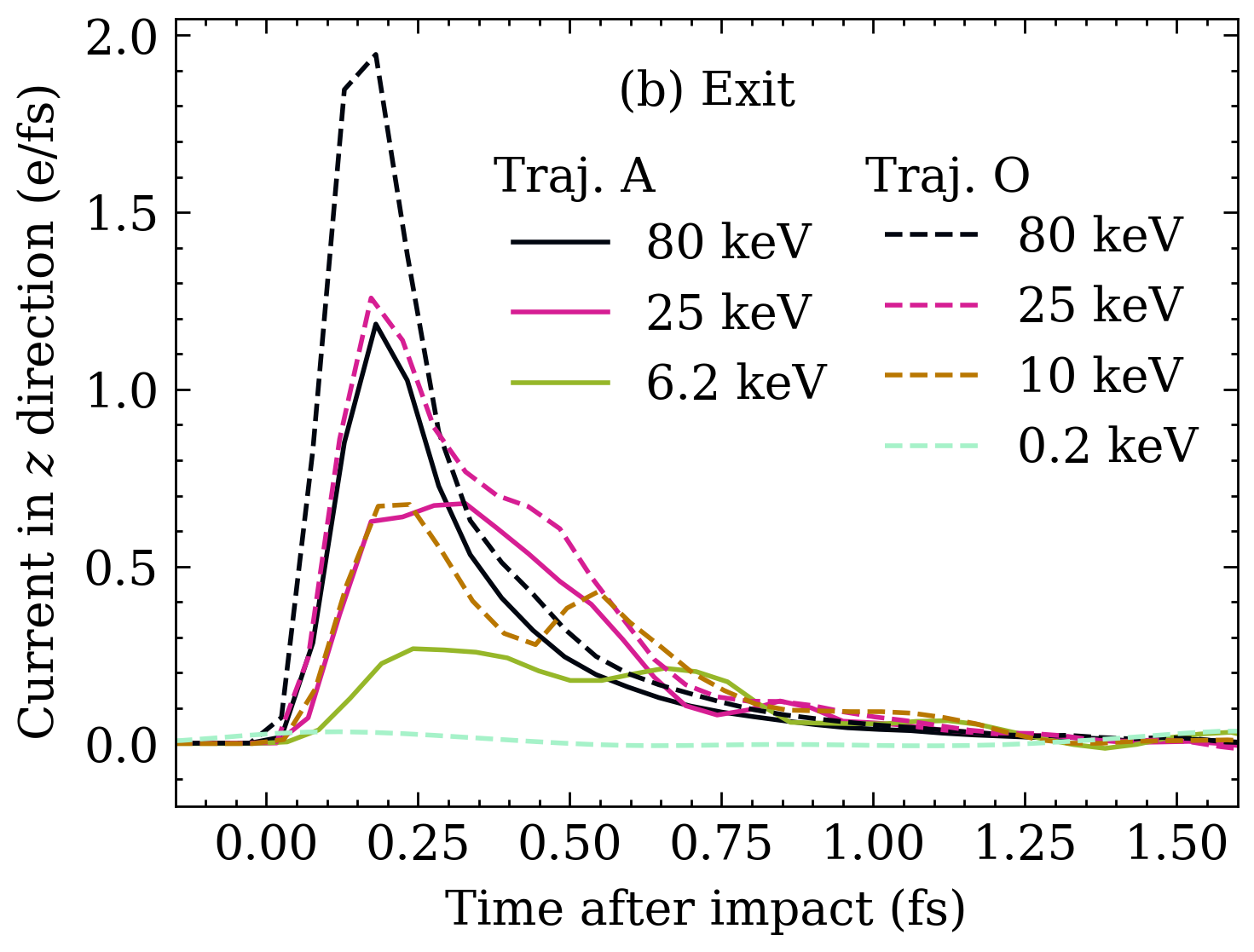}
\caption{\label{gchar}
Simulated secondary electron current on the (a) entrance and (b) exit sides for proton-irradiated monolayer graphene (see ``Methods'' and the SI for 
computational details).
The proton is exactly in the graphene plane at ``impact''.
}
\end{figure}

Our simulations show that the peak intensity of the secondary electron current in Fig.\ \ref{gchar} follows the same trend with proton kinetic energy as the secondary electron yield reported in Ref.\ \cite{kononov_anomalous_2021}:
The overall peak intensity is higher at the exit side than the entrance side, with a maximum at around 10 keV for the entrance side and 80 keV for the exit side (see details in Sec.\ \ref{sec:sitotalemit} in the SI).
Experimental ratios of the exit- and entrance-side yields range from 1 to 1.5 in a comparable velocity range for hydrogen-irradiated thin carbon foils \cite{ritzau_electron_1998}, slightly below our computed yield ratios of 1\,--\,2.75.

The difference between entrance and exit sides for peak intensity (see Fig.~\ref{gchar}) and secondary electron yield (see Fig.~\ref{fig:seccom}) suggests that multiple emission mechanisms contribute. 
Potential emission of secondary electrons occurs due to potential energy released as the projectile neutralizes its charge state in the target.
The magnitude of potential emission near exit and entrance sides depends on the projectile's ionization energy and the target material's work function and Fermi velocity~\cite{kishinevsky_estimation_1973,baragiola_electron_1979}.
Although potential emission is more commonly discussed for highly charged ions~\cite{schwestka_charge-exchange-driven_2019, niggas_ion-induced_2022}, the potential energy stored in the proton (13.6 eV) is almost three times the work function of graphene (4.56 eV)~\cite{yu_tuning_2009,yan_determination_2012}, which exceeds the minimum requirement for Auger-Meitner neutralization as a mechanism for potential emission \cite{kishinevsky_estimation_1973, baragiola_electron_1979}.  
Secondary electrons due to kinetic emission preferentially travel in the same direction as the proton due to momentum conservation, causing more emission on the exit-side.
On the other hand, longer time-scale emission mechanisms involving decaying excitations within the material are expected to be symmetric with respect to emission side.

To explore potential emission effects, we study a proton velocity of 0.1 at.\,u.\ (0.2 keV), which is below the kinetic emission threshold estimated at 0.13\,--\,0.19 at.\,u.\ (0.42\,--\,0.9 keV)~\cite{kononov_anomalous_2021}.
Our data in Fig.\ \ref{gchar}(a) shows pre-impact emission for this case, where potential emission is expected to be the primary contribution. 
For this proton kinetic energy, the secondary electron yield for entrance and exit sides is around 12\,\% of the maximum yield in our simulations (see Fig.~\ref{fig:seccom}).
The expression derived in Ref.~\onlinecite{kishinevsky_estimation_1973}, which shows fair agreement with experiment~\cite{lakits_threshold_1990, baragiola_electron_1979}, predicts a potential emission yield around 0.09 (see Sec.~\ref{sec:pee} in the SI), a value comparable to our predictions of around 0.027 (0.065) on the entrance (exit) side. 
The yield ratio of about 2.5 in our simulations indicates asymmetric secondary electron emission also in the potential emission regime, albeit with a very small secondary electron current (see Fig.\ \ref{gchar}).
Additionally, electron capture is most prominent for slow protons~\cite{kononov_anomalous_2021,zhao2014comparison}, further suggesting that potential emission plays a role in this regime.
However, the relatively small potential energy stored in the proton leads to a small corresponding secondary electron yield  
compared to slow but highly charged ions~\cite{schwestka_charge-exchange-driven_2019,niggas_ion-induced_2022}.

\begin{figure}
\includegraphics[width=0.95\columnwidth]{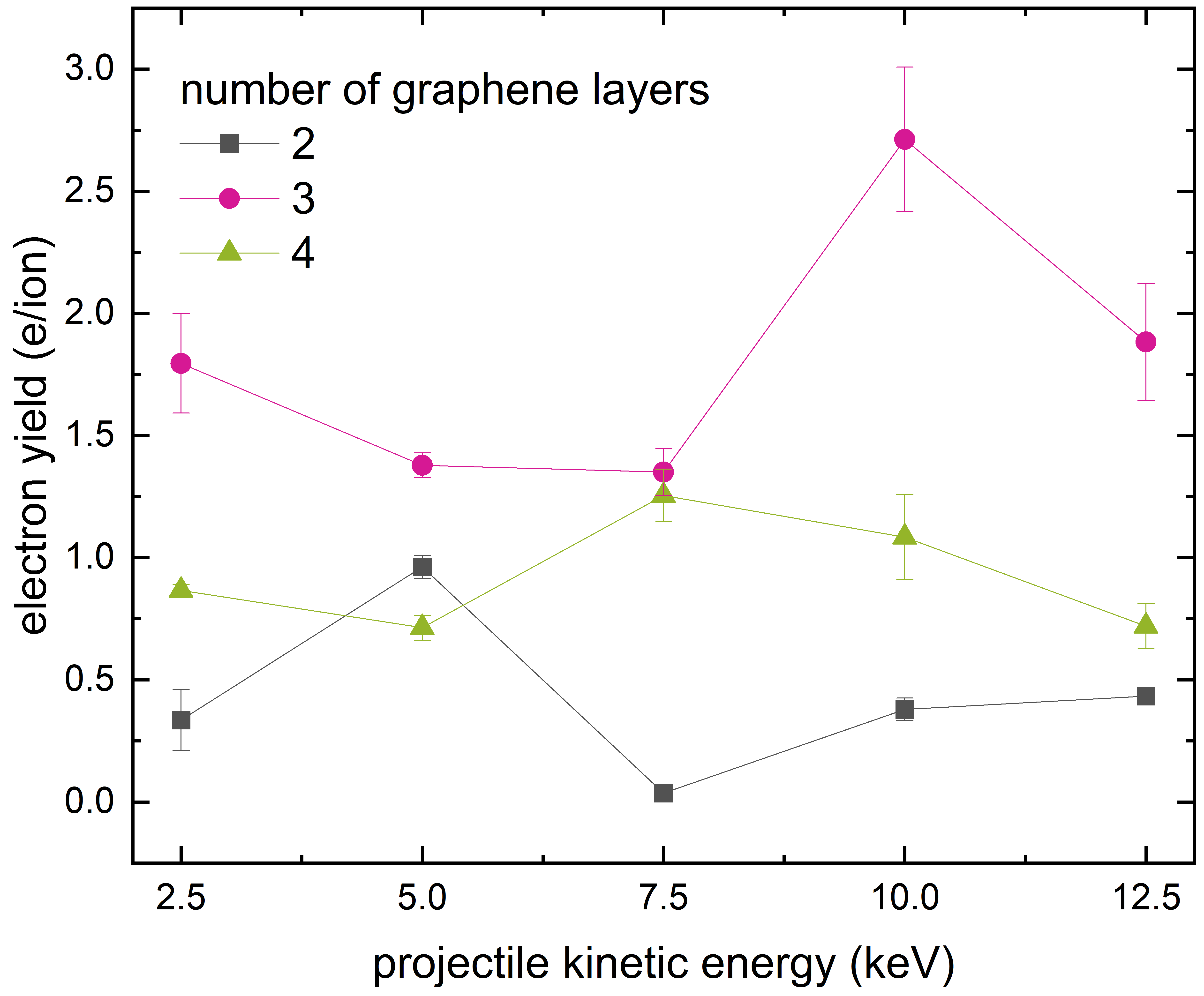}
\caption{\label{fig:PIPS_Yield}
Experimental electron yield $\gamma$ from graphene samples with different layer numbers as a function of the projectile ($\mathrm{Ar^+}$) energy.
The yield was calculated from the electron spectrum measured with the PIPS detector by fitting a Furry distribution to the data. 
Lines are a visual guide for the eye.
}
\end{figure}

To support and complement our simulations, we also investigated secondary electron emission experimentally (see ``Methods'' and the SI for experimental details).
Figure \ref{fig:PIPS_Yield} shows the dependence of the measured electron yield on the number of graphene layers for Ar$^{+}$ ions with different kinetic energies.
We find that there is no strong dependence on the kinetic energy of the projectile. 
Instead, 
the observed yield mainly varies with the number of graphene layers:
We find yields of about 0.4, 1.8, and 0.7 electrons per ion for two-layer, three-layer, and four-layer graphene, respectively.

The velocity of the 12.5 keV Ar$^{+}$ ion is about 0.1 at.\,u., comparable to the velocity of the 0.2 keV proton simulation, for which
potential emission mechanisms are expected to dominate.
For this velocity, the larger experimental yields (see Fig.\ \ref{fig:PIPS_Yield}) compared to our simulation predictions (see Fig.\ \ref{fig:seccom}) can be understood in terms of the model of Ref.\ \onlinecite{kishinevsky_estimation_1973}. 
The ionization energy of argon is 15.8 eV~\cite{weitzel_zeke-pepico_1994}, slightly higher than the proton (13.6 eV). 
The theoretical model~\cite{kishinevsky_estimation_1973} (see Sec.~\ref{sec:pee} in the SI) predicts a potential emission yield of around 0.18 for Ar$^{+}$ irradiating monolayer graphene, on the same order as most of our experimental results for two-layer graphene.
The factor of $\sim 2$ discrepancy between the experimental yields and model prediction may be attributable to the different graphene thicknesses. 

Furthermore, our results in Fig.\ \ref{gchar} show that for protons with kinetic energies above the kinetic emission threshold, secondary electron emission from the entrance and exit sides ends after only about 1\,--\,1.5 fs. 
A comparable electron emission duration was also reported for graphene under irradiation by highly charged ions~\cite{niggas_ion-induced_2022}.
This short time scale indicates that delayed secondary electron emission, resulting from the relaxation of the energy deposited during ion impact, is not observed in our simulations.
In particular, while the dipole moment of the graphene after proton impact shows distinct plasmonic oscillations (see Fig.~\ref{fig:dip} in the SI), our real-time profile of the secondary electron emission does not show indications of plasmon decay.
This is not surprising, since the intrinsic plasmon lifetime~\cite{ni_fundamental_2018} of about 12 ps is much longer than the timescale of our simulations.

\begin{figure}
\includegraphics[width=0.95\columnwidth]{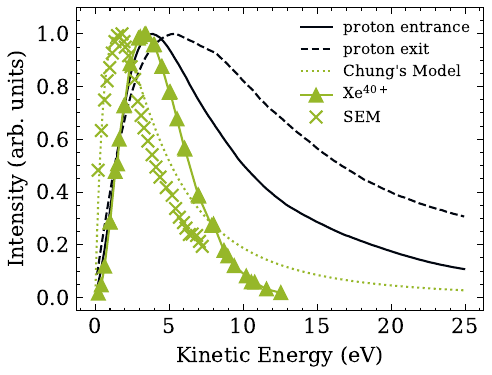}
\caption{\label{fig:litcom}
Comparison of our simulated kinetic energy spectra (centroid trajectory, kinetic energy of 80 keV in Fig.\ \ref{fig:keall}) with a measured spectrum for $\mathrm{Xe^{40+}}$ ions impacting graphene~\cite{schwestka_charge-exchange-driven_2019}, a measured spectrum under electron irradiation in scanning electron microscopy (SEM) experiments~\cite{zhou_quantitative_2016}, and an analytical expression derived by Chung~\cite{chung_simple_1974} (see details in the text).
The $y$ axis is normalized to the same peak intensity.
}
\end{figure}

Next, we analyze the kinetic energy spectrum of the emitted electrons as a probe of the local electronic structure of the target and of the local escape probability of electrons near the projectile impact point.
Our simulation results in Fig.\ \ref{fig:litcom} show a characteristic peak around 3 eV for entrance side and around 5 eV for exit side.
Since the peak positions do not strongly depend on impact point and proton kinetic energy (see details in Sec.~\ref{sec:SIKEspectra} of the SI), we conclude that this peak is predominantly governed by intrinsic material properties of the graphene target, such as the mean free path of excited electrons and the work function.
The general shape of our kinetic energy spectra in Fig.\ \ref{fig:litcom} agrees well with experimental data and with the analytical expression developed by Chung~\cite{chung_simple_1974}, that 
relates the position of the secondary electron peak to the work function of the material.
For graphene, with a work function of approximately 4.56 eV~\cite{yu_tuning_2009, yan_determination_2012}, it predicts a peak of position of $4.56/3=1.52$ eV, which is about 2 eV below our simulation results.
The scanning electron microscopy (SEM) data \cite{zhou_quantitative_2016} is obtained in reflection geometry and should thus be compared to our entrance side simulations.
For this data, the Chung model captures peak position and shape very accurately.

In Fig.\ \ref{fig:litcom} we also include data from highly charged Xe$^{40+}$ irradiation \cite{schwestka_charge-exchange-driven_2019,niggas_ion-induced_2022}, which is measured in forward direction and should be compared to our exit-side simulations.
For ion irradiation experiments as well as our proton simulation, the kinetic energy peak appears shifted to higher kinetic energies by about 2 eV relative to Chung's model.
Although there is a modified version of this model \cite{petrov_secondary_2011,zhou_quantitative_2016} that extends it to helium irradiation, these modifications rely on fitting the energy distribution of the excited electrons to the experimental data. 
Those parameters might then be material dependent, and hence, we do not compare with the modified Chung model in this work.
In addition, the constant value of the work function used in Chung's model will be modified by the radiation-induced electron dynamics~\cite{niggas_ion-induced_2022}. 
Other processes, such as attraction from the positively charged incident ion before impacting the target, might lead to further deviations for ion-induced emission since Chung's model is derived for electron irradiation. 

Finally, we investigate how the local, pre-excited target material responds to external ion irradiation and how this affects intensity and duration of secondary electron pulses.
Non-equilibrium excited electrons, e.g.\ due to proton irradiation, will thermalize within a ps or less through electron-electron scattering and then heat the lattice through electron-phonon scattering.
This timescale is comparable to the time interval between successive ion impacts in modern ion sources.
In addition, our RT-TDDFT simulations revealed that the duration of the secondary electron pulse is only around 1.5 fs (see Fig.\ \ref{gchar}).
This is a fast time scale compared to thermalization, suggesting that the intensity of secondary electron emission can serve as a probe of the thermalization processes in the target material.
To separate the effects of elevated electron and lattice temperatures, we investigated proton irradiation of a cold lattice with elevated electronic temperature and an elevated lattice temperature with cold electrons.
We use electron and lattice temperatures from previous works based on two temperature models \cite{duvenbeck_computer_2004,zarkadoula_electronic_2014,fieret_fundamentals_2005}.

\begin{figure}
\includegraphics[width=0.95\columnwidth]{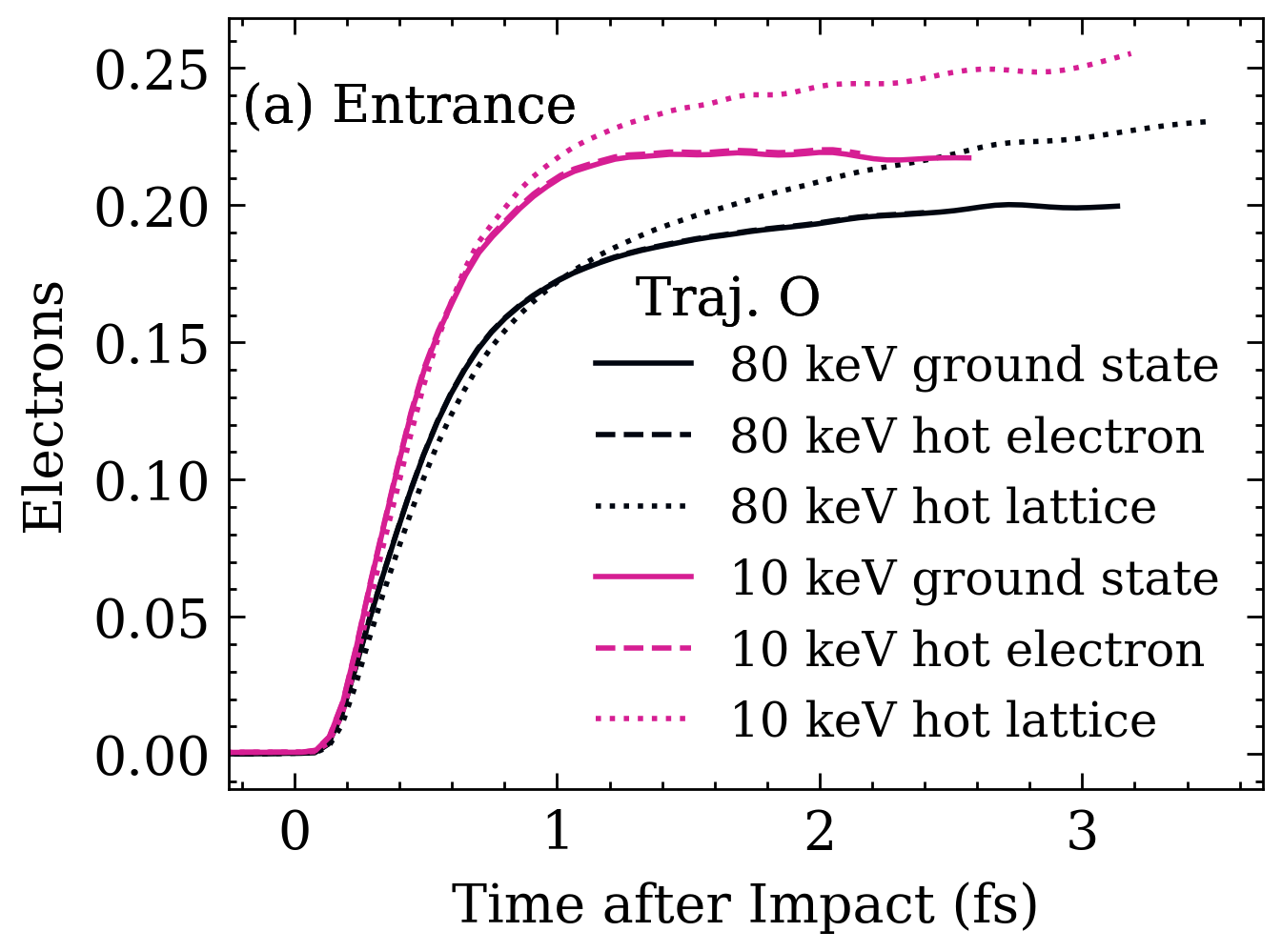}\\
\includegraphics[width=0.95\columnwidth]{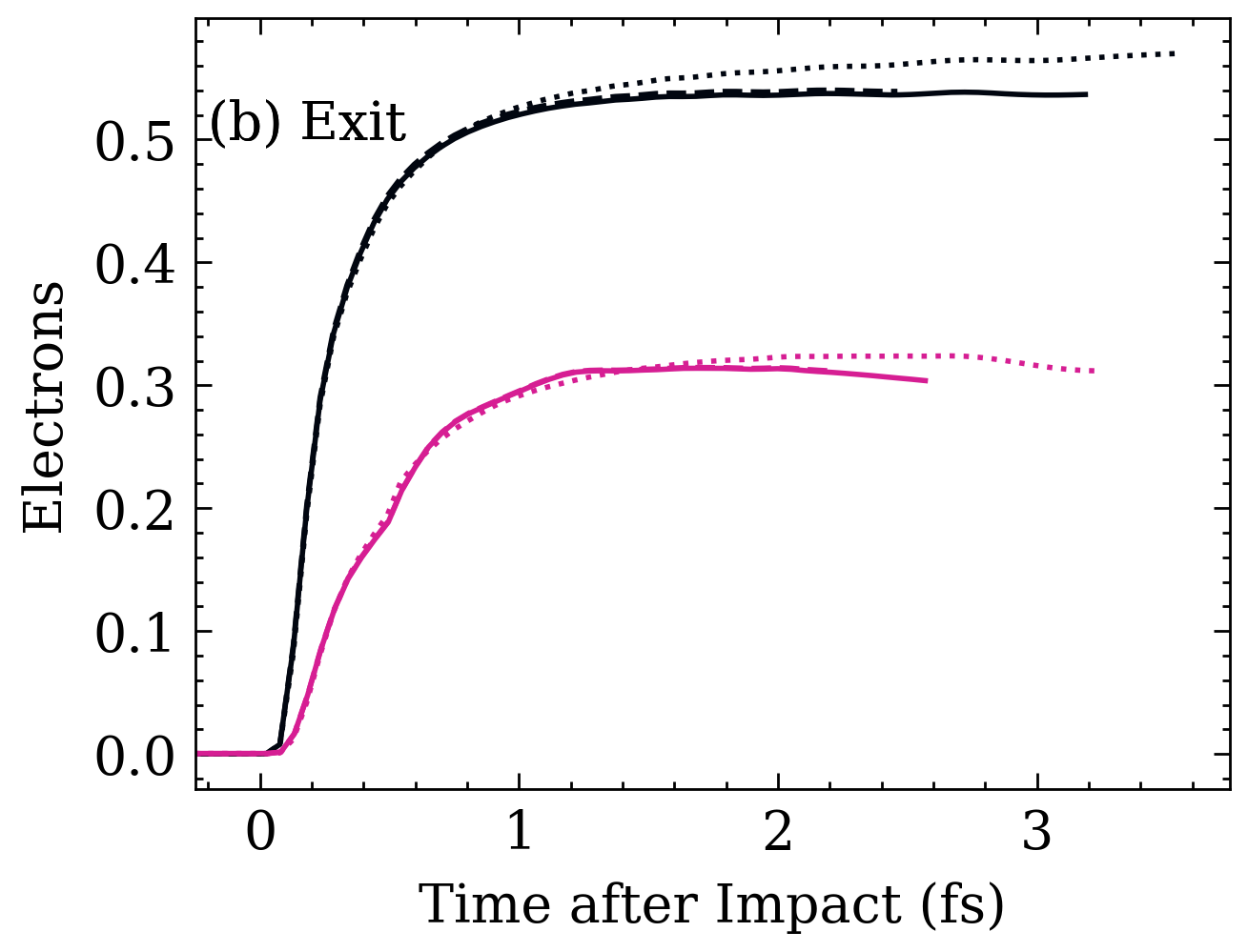}
\caption{
\label{fig:Tcom}
Number of secondary electrons, computed as described in ``Methods'',
at elevated electron temperature of $T_e = 10,000$ K or elevated lattice temperature of $T_i = 1,000$ K, compared to electron emission from ground-state graphene at $T_e = T_i = 0$ K.
The proton is exactly in the graphene plane at ``impact''.
}
\end{figure}

First, we compute the secondary electron yield for an electronic temperature of 10,000 K (see Fig.\ \ref{fig:Tcom}) which corresponds to $2.76\times 10^{20}\,\mathrm{cm^{-3}}$ thermally excited electrons and holes. 
This electronic temperature is exaggeratedly high for ion irradiation after 1 ps,
but achievable within the first 100 fs after the initial impact, as predicted in Refs.~\cite{duvenbeck_computer_2004,zarkadoula_electronic_2014}.
Nevertheless, our simulations show that the secondary electron yield is very similar to that from the ground state graphene (see Fig.\ \ref{fig:Tcom}), and the effect of even such a high electron temperature is minor. 

On the contrary, our data shows that a lattice temperature of 1,000 K significantly increases the secondary electron yield by about 10\,--\,13\,\% (3\,--\,5\,\%) on the entrance (exit) side relative to the ground state.
This enhanced emission also occurs for the channeling trajectory (see Fig.~\ref{fig:TAOcom}).
Such a lattice temperature was reported after about 400 fs from molecular dynamics simulations of ion-irradiated Fe \cite{zarkadoula_electronic_2014}. 
Besides secondary electron yield, we also simulate the kinetic energy spectrum at this elevated lattice temperature (see Fig.~\ref{fig:ClatticeKE}).
The peak position and shape of the spectrum resemble those of the ground state and we predict only enhanced secondary electron yield as an indicator of lattice temperature.

\begin{figure}
\includegraphics[width=0.95\columnwidth]{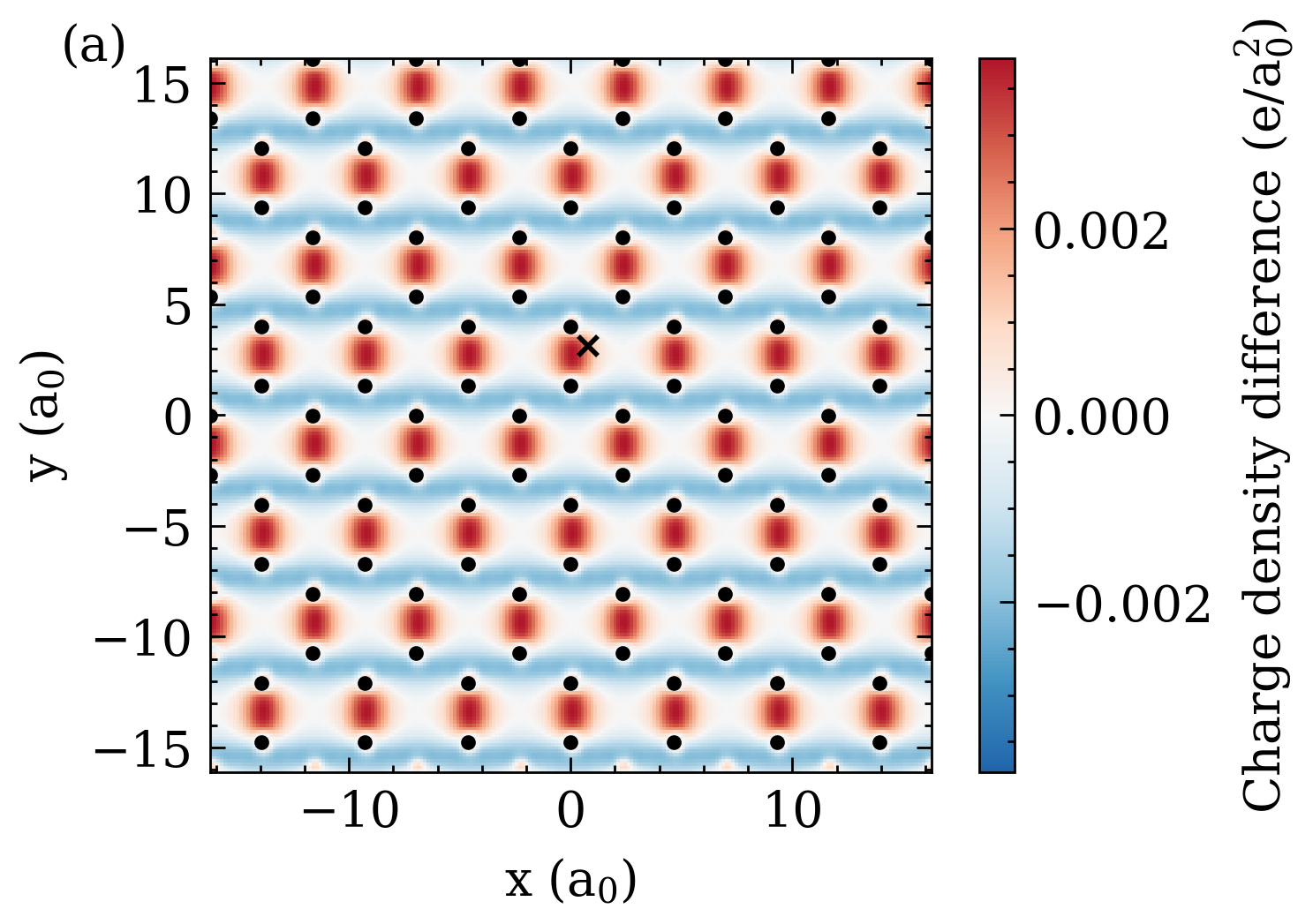}\\
\includegraphics[width=0.95\columnwidth]{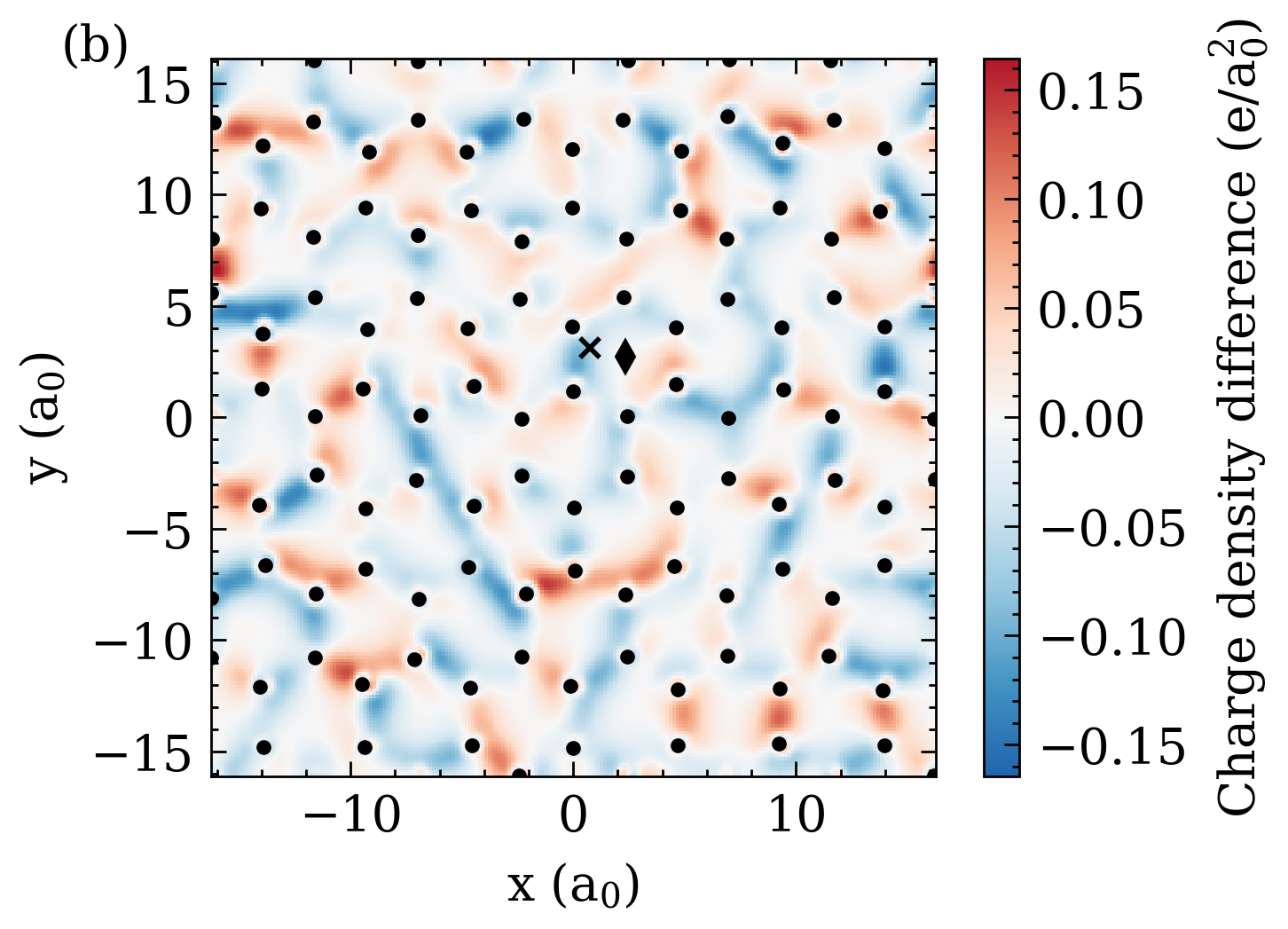}
\caption{\label{fig:gsrhoslab}
Comparison of the planar integrated charge density difference between the elevated (a) electron ($T_e=10,000$ K) or (b) lattice ($T_i=1,000$ K) temperature and ground-state graphene for the initial state before real-time propagation.
Black circles denote atomic positions in the hot graphene.
The black diamond (X) denotes the proton impact point of the channeling (centroid) trajectory.
}
\end{figure}

To explain the modified secondary electron yield we analyze the electron density distribution in Fig.\ \ref{fig:gsrhoslab}.
Even though an electronic temperature of 10,000 K is very high, it results in only a small charge density difference compared to the ground-state electron density of graphene.
Conversely, the electron density difference at an elevated lattice temperature is about two orders of magnitude higher, significantly modifying the interaction between projectile and target material.
We note that the enhanced electron emission does not just depend on the charge density difference exactly at the impact point, since the long-range Coulomb interaction decays as $1/r$.

In summary, through a combination of experiments and real-time time-dependent density functional theory, we quantitatively investigate the non-equilibrium dynamics of secondary electron emission in hot and cold proton-irradiated graphene.
We show that the secondary electron pulse concludes within about 1.5 fs, much faster than sub-ps thermalization processes of non-equilibrium electrons. 
While the secondary electron yield depends sensitively on the trajectory and kinetic energy of the projectile, the shape of the secondary electron pulse and the kinetic energy spectra agree well among different impact parameters. 
Our newly derived method for simulating kinetic energy spectra agrees well with literature reports for electron and ion projectiles, with peak positions varying by about 1\,--\,2 eV. 
Additionally, pre-excited graphene during successive ion irradiation shows enhanced electron emission due to electron density perturbations at high lattice temperatures.
Further experimental work is needed to verify predicted electron yields at non-vanishing lattice or electron temperatures. 
Nevertheless, we conclude that the characteristics of the secondary electron yield as a probe pick up on \emph{lattice} thermalization, and not electron thermalization, in the graphene e.g.\ in highly promising, recently developed ion-pump-probe experimental techniques. 

\section*{\label{sec:method}Methods}

We perform real-time time-dependent density functional theory simulations starting from converged ground-state single-particle Kohn-Sham (KS) states computed using density functional theory.
These are propagated in real time through the time-dependent KS equations 
\begin{equation}
\label{eq:ks}
i\frac{\partial}{\partial t} \phi_j(\mathbf{r},t)=\left(-\frac{\nabla^2}{2}+V_{\mathrm{KS}}(\mathbf{r},t)\right)\phi_j(\mathbf{r},t),
\end{equation}
using the enforced time-reversal symmetry (ETRS) method with a time step of one attosecond.
Here, $\phi_j$ are the single-particle KS orbitals evolving in a time-dependent effective potential $V_{\mathrm{KS}}$ which is a functional of the electron density $n(\mathbf{\mathrm{r}},t)$.
Kohn-Sham states are represented using a plane-wave basis with a cutoff kinetic energy of 50 Hartree.
Exchange and correlation effects are treated using the adiabatic local density approximation~\cite{zangwill_resonant_1980, zangwill_resonant_1981}.
The electron-ion interaction is described by norm-conserving Hamann-Schl\"uter-Chiang-Vanderbilt pseudo-potentials~\cite{vanderbilt_optimally_1985}.
The large simulation cell with 112 carbon atoms and 100 $\mathrm{a_0}$ of vacuum, visualized in Fig.\ \ref{fig:cell} along with the channeling (A) and centroid (O) proton trajectories, allows
using only the $\Gamma$ point for Brillouin zone sampling.
All these parameters are consistent with those used previously \cite{kononov_anomalous_2021, kononov_first-principles_2022, kononov_electron_2022}.
We use absorbing boundary conditions \cite{kononov_electron_2022} as implemented in the Qb@ll code~\cite{draeger_qbox_2017,schleife_quantum_2014,draeger_massively_2017}, to avoid the unphysical re-entering of emitted electrons into a periodic image of the simulation cell (see details in the SI).
We represent an elevated electronic temperature using a Fermi-Dirac distribution of KS occupation numbers within Mermin DFT~\cite{mermin_thermal_1965}.
An elevated lattice temperature is represented using atomic displacements that we create using the approach described in Ref.\ \cite{aberg_contributions_2013} and the SI.

The number of emitted electrons in the entrance- and exit-side vacuum are determined by integrating the current density passing through an artificial detector surface, placed 10.5 $a_0$ away from the graphene.
The number of electrons captured by the proton is determined by integrating the charge density over a sphere with a radius of 5 $a_0$ around the proton.
Except for computing kinetic energy spectra in Figs.\ \ref{fig:litcom} and \ref{fig:keall}, we subtract these captured electrons from the total emitted electrons in this work, because they would not be measured in the experiment as secondary electrons.

To obtain the kinetic energy distribution at time $t$ (in atomic units where $m_e$=1), we devise a new method based on the number of electrons $n(\mathbf{r},t)\,d\mathbf{r}$ in the entire vacuum with a specific kinetic energy $v^2(\mathbf{r}, t)/2$:
\begin{equation}
\label{eq:ke hist}
P(\mathrm{KE},t)= \int_{V_{\mathrm{vac}}} n(\mathbf{r}, t) \, \delta\left(\mathrm{KE}-\frac{v^2(\mathbf{r}, t)}{2}\right) \, d\mathbf{r},
\end{equation}
where $d\mathbf{r}$ is the volume attributed to each point of the real-space grid that discretizes the electron density and $V_\mathrm{vac}$ is the vacuum volume on each side of the graphene between the detector surface (10.5 $a_0$ from graphene) and the absorbing potential boundary (40 $a_0$ from graphene),
and $v(\mathbf{r}, t) = j(\mathbf{r},t)/n(\mathbf{r},t)$ is the electron velocity distribution given by the current density $j$ and electron density $n$.
At each time step, we histogram the number of electrons corresponding to the specific kinetic energy and we integrate these instantaneous spectra over time to include all emitted electrons.
We discuss the challenges and benefits of this new method and compare to the time-of-flight method in Sec.\ \ref{sec:ke spectrum} of the SI.

Experimentally we investigated the number of electrons emitted from a free-standing graphene target under singly charged ion bombardment.
The experimental setup to investigate the electron emission (see Fig.\ \ref{fig:PIPS} of the SI) is based on a design originally developed by the Aumayr group \cite{Aumayr_PIPS_1991, Schrempf_PIPS_2013}. 
The experiment operates in transmission geometry with free-standing graphene targets consisting of multiple layers (see sample preparation details in Ref.\ \onlinecite{Kalkhoff_ultralarge_2023}).
A singly charged argon ion produced by a rare gas ion source (Atomica Duoplasmatron) impacts the graphene target with kinetic energies of 2.5\,--\,12.5~keV, leading to electron emission. 
Notably, no stripping effects are expected as the rather slow singly-charged argon ions traverse the target, and thus the experiments are comparable to our proton-irradiation simulations.
A proton with the same velocity as an argon ion has a 40 times smaller kinetic energy.
Hence, we expect that irradiation with $\sim 6.2$ keV protons, as simulated in this work, should produce comparable charge and velocity conditions for secondary electron emission as in Ref.\ \onlinecite{wittkower_charge_1973}. 
The emitted electrons are accelerated with a high voltage of 25~kV towards a passivated implanted planar silicon (PIPS, Mirion Technologies) detector.
The resulting spectra are fit with a Furry distribution\ \cite{Furry_Distr_1937,Dietz_FurryDistr_2008}, which is a Polya distribution where $b=1$, to evaluate the electron yield.

\begin{acknowledgments}
We thank Zhihao Jiang, Richard Arthur Wilhelm, and Anna Niggas for valuable scientific discussions.
This material is based upon work supported by the National Science Foundation under Grant Nos.\ OAC-2209857 and OAC-1740219, and by the Deutsche Forschungsgemeinschaft (DFG, German Research Foundation) Project No.\ 278162697-SFB 1242.
A.S.\ acknowledges support by the Mercator Fellow Program of SFB 1242.
A.K.\ was partially supported by the US Department of Energy Science Campaign 1 and Sandia National Laboratories' Laboratory Directed Research and Development (LDRD) Project No.\ 233196.
An award of computer time was provided by the Innovative and Novel Computational Impact on Theory and Experiment (INCITE) program.
This research used resources of the Argonne Leadership Computing Facility, which is a DOE Office of Science User Facility supported under Contract DE-AC02-06CH11357.
This work made use of the Illinois Campus Cluster, a computing resource that is operated by the Illinois Campus Cluster Program (ICCP) in conjunction with the National Center for Supercomputing Applications (NCSA) and which is supported by funds from the University of Illinois at Urbana-Champaign.
This article has been co-authored by an employee of National Technology \& Engineering Solutions of Sandia, LLC under Contract No. DE-NA0003525 with the U.S. Department of Energy (DOE). The authors own all right, title and interest in and to the article and are solely responsible for its contents. The United States Government retains and the publisher, by accepting the article for publication, acknowledges that the United States Government retains a non-exclusive, paid-up, irrevocable, world-wide license to publish or reproduce the published form of this article or allow others to do so, for United States Government purposes. The DOE will provide public access to these results of federally sponsored research in accordance with the DOE Public Access Plan \url{https://www.energy.gov/downloads/doe-public-access-plan}.
\end{acknowledgments}

\bibliography{lib}

\newpage

\beginsupplement

\section*{Supporting Information}

\subsection{Experimental setup}

\begin{figure}[b]
\includegraphics[width=1.0\columnwidth]{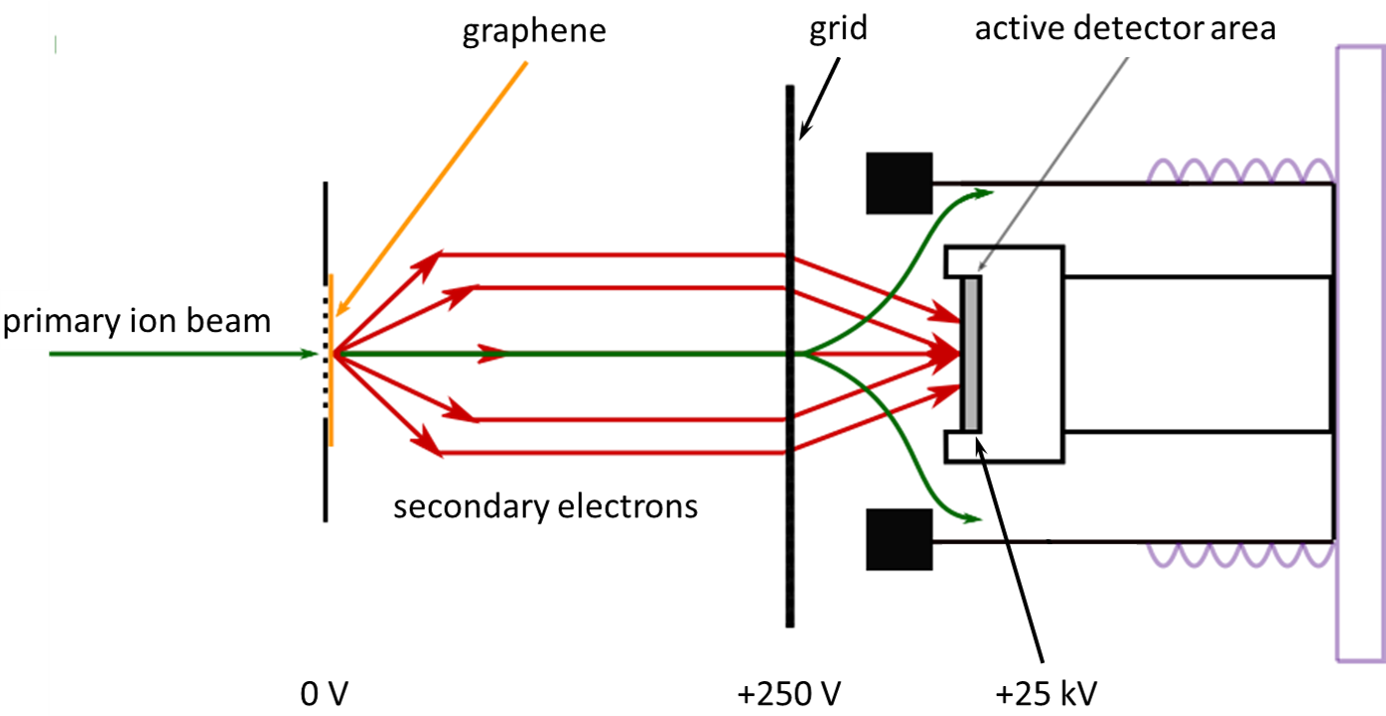}
\caption{
\label{fig:PIPS}
Schematic of the experimental set-up. The ion beam is directed onto a free-standing graphene film in transmission geometry. The emitted electrons are accelerated via increasing voltages towards the PIPS detector. The potential of the detector is high enough to deflect the transmitted ions.
}
\end{figure}

The experimental setup for investigating electron emission is depicted in Fig.\ \ref{fig:PIPS}.
It is based on a design originally developed by the Aumayr group that is described in great detail in Refs.\ \cite{Aumayr_PIPS_1991, Schrempf_PIPS_2013}. 
The experiment operates in transmission geometry with free-standing graphene targets consisting of multiple layers (2-4) (see Ref.\ \cite{Kalkhoff_ultralarge_2023} for details of the sample preparation).
A singly charged argon ion produced by a rare gas ion source (Atomica Duoplasmatron) impacts the graphene target with kinetic energies between 2.5~keV and 12.5~keV, leading to electron emission. 
It is important to note that due to the rather low velocity of the projectile, no stripping effects are expected while it traverses the target, and the singly charged argon ions are thus comparable to the proton projectiles in our simulations.
Also, experiments with up to approximately 100 keV argon projectiles impacting thin carbon foils found that the average equilibrium charge state is less than one \cite{wittkower_charge_1973}.
A proton with the same velocity as a 100 keV argon 
ion has a kinetic energy of about 2.5 keV.
Hence, we expect that irradiation with $\sim 6.2$ keV protons, as simulated in this work, should produce comparable charge and velocity conditions for secondary electron emission as in Ref.\ \cite{wittkower_charge_1973}. 

The emitted electrons are accelerated with a high voltage of 25~kV towards a passivated implanted planar silicon (PIPS, Mirion Technologies) detector.
In principle, this is an energy-resolved detection device with a resolution of approximately 3~eV defined by the direct band gap in the doped silicon chip. 
Therefore, a single electron reaching the PIPS detector with a kinetic energy of 25~keV will produce about 8000 electron-hole pairs, resulting in a well-defined current pulse at the output of the detector.
From this, we precisely determine the number of electrons emitted from the graphene target per ion impact, because $n$ electrons will produce exactly $n$ times the pulse of a single electron.

The experiment is not time-resolved and therefore not capable of directly following the dynamics in graphene triggered by the ion impact.
Nevertheless, it provides crucial experimental information about the total number of emitted electrons per ion impact from graphene, supporting simulation results and vice versa.

\subsection{Simulation cell and absorbing boundaries}
\label{sec:simulationcell}

\begin{figure}[b]
\includegraphics[width=0.95\columnwidth]{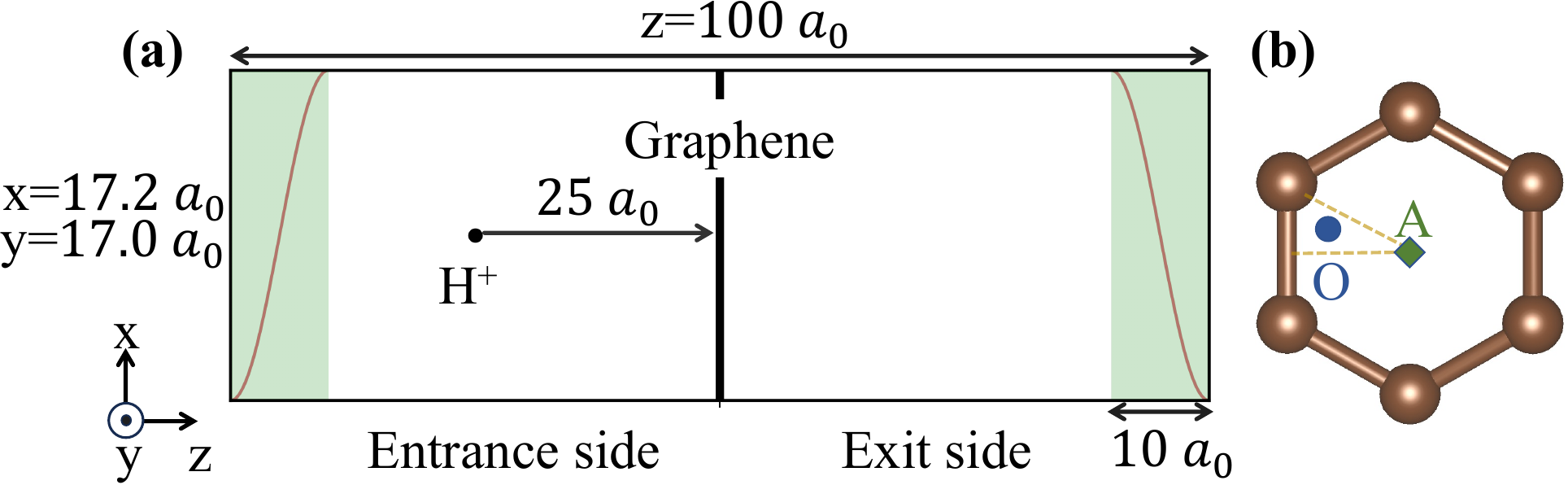}
\caption{\label{fig:cell}
(a) Schematic of the simulation cell used in this work.
The proton is initially placed 25 $a_0$ away from the graphene and travels normally to the graphene plane with fixed velocity.
The green region is where the complex absorbing potential (CAP, brown curves) is $< 0$, see Eq.\,\eqref{eqn:cap}.
(b) Both the channeling (A) and centroid (O) impact points of the proton in monolayer graphene are studied in this work.
}
\end{figure}

The simulation cell used in this work is illustrated in Fig.\ \ref{fig:cell}.
Real-time propagation begins with the proton placed 25 $\mathrm{a_0}$ away before it traverses the graphene with a constant velocity along a trajectory normal to the graphene plane on either a channeling (A) or centroid (O) trajectory (see Fig.\ \ref{fig:cell}b).
Carbon nuclei are fixed since the femtosecond time scale of proton impact is too short for them to move appreciably, and our longest simulation is only about 12 fs.
This finite simulation length only affects secondary electron emission occuring on longer time scales, e.g.\ due to plasmon decay or carbon ion recoil.

\begin{figure}
\includegraphics[width=0.95\columnwidth]{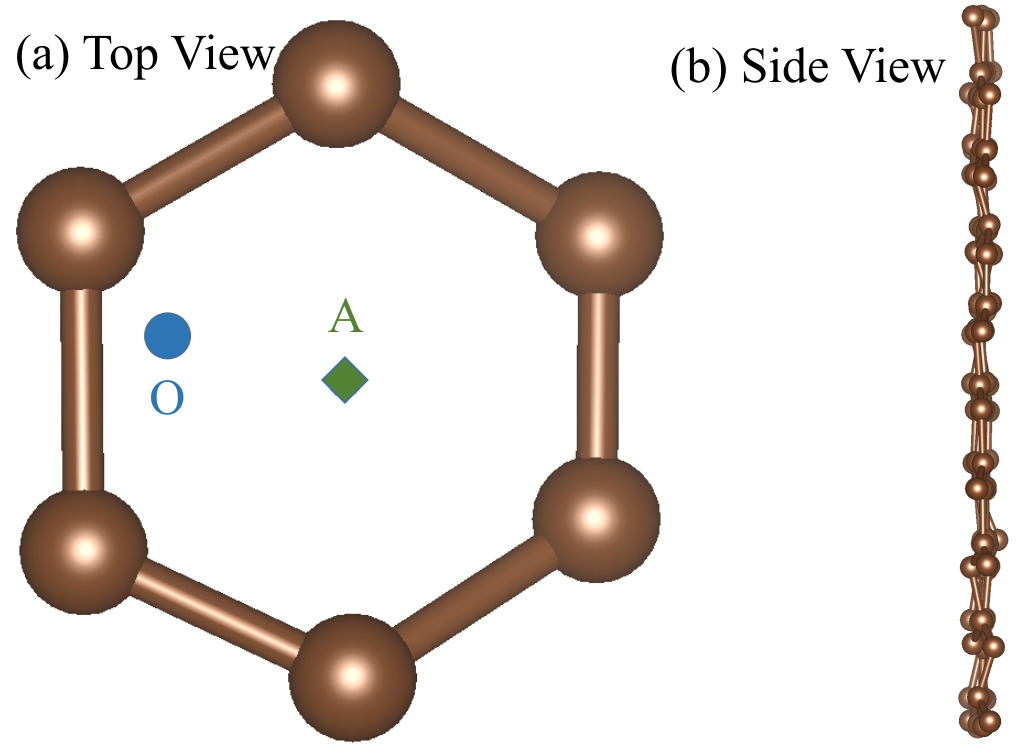}
\caption{\label{fig:lattT}
Carbon atoms in graphene near proton impact points at $T_i=1,000$ K in (a) a top view and (b) a side view. Due to the distortion of the atomic positions at 1,000 K lattice temperature, the impact point of the channeling and centroid trajectories are not exactly identical compared to those at 0 K (see Fig.\ \ref{fig:cell}b).
}
\end{figure}

In this work, we perform simulations of secondary electron emission from ground-state graphene and compare it to graphene with elevated lattice and electronic temperatures.
We represent an elevated electronic temperature using a Fermi-Dirac distribution of KS occupation numbers within Mermin DFT~\cite{mermin_thermal_1965}.
An elevated lattice temperature is represented using atomic displacements that we create using the approach described in Ref.\ \cite{aberg_contributions_2013}.
In particular, we superimpose harmonic phonon modes with random phases and amplitudes according to classical statistics.
The mean-square amplitude of the lattice vibrations follows from
\begin{equation}
\left< \left | Q_j \right | \right> = \frac{k_BT_i}{\omega^2_j},
\end{equation}
where $\omega_j$ is the frequency of the $j$-th phonon mode, $Q_j$ are the corresponding normal coordinates~\cite{Eckold2003}, and $T_i$ is the lattice temperature.  
All phonon eigenmodes are calculated via the finite-difference method as implemented in the Vienna \emph{Ab-initio} Simulation Package (\texttt{VASP})~\cite{kresse_efficient_1996,kresse_ultrasoft_1999}.
The projector-augmented wave (PAW) method\cite{blochl_projector_1994} is used to describe the electron-ion interaction. We use a relaxed 112-atom supercell with vacuum length reduced to 39.3 $a_0$,
a $4\times4\times1$ \mbox{$\mathbf{k}$-point} grid, and 530 eV as the kinetic energy cutoff of the plane-wave basis.
We then use a single snapshot resulting from this method to represent the atomic geometries in our Qb@ll RT-TDDFT simulation of graphene with a large lattice temperature for multiple proton trajectories.
The distortion of the atomic positions compared to a cold lattice alters the exact positions of the proton impact points relative to the carbon ions, which may influence electron emission.
The structure and impact points at finite lattice temperature are shown in Fig.\ \ref{fig:lattT}.

As the proton impacts graphene, secondary electrons are emitted from both sides and propagate outwards.
The complex absorbing potential \cite{de_giovannini_modeling_2015, kononov_electron_2022} used in this work is represented as
\begin{equation}
\label{eqn:cap}
V_{\mathrm{CAP}}(z)=-i\cdot W\cdot \sin^2\left(\frac{(z-z_\mathrm{s})\cdot \pi}{2\cdot dz}\right),
\end{equation}
for $z_\mathrm{s}<z<z_\mathrm{s}+2dz$, where $z_\mathrm{s}$ is the front boundary of the CAP, $dz$ is its half width, and the graphene is at $z=0$. 
In our simulations, we set $z_\mathrm{s}$=$40~a_0$, $dz$=$10~a_0$, and $W$=$15$ Ha.
To avoid artificial re-entering of the proton due to periodic boundary conditions, we remove it once it reaches the cell boundary. 
An explicit comparison between the secondary electron yield computed in this work (with absorbing boundary conditions) and previous results (without absorbing boundary conditions)~\cite{kononov_anomalous_2021} is shown in Fig.~\ref{fig:seccom}.

\subsection{\label{sec:sitotalemit}Total emitted electrons}

\begin{figure}
\includegraphics[width=0.95\columnwidth]{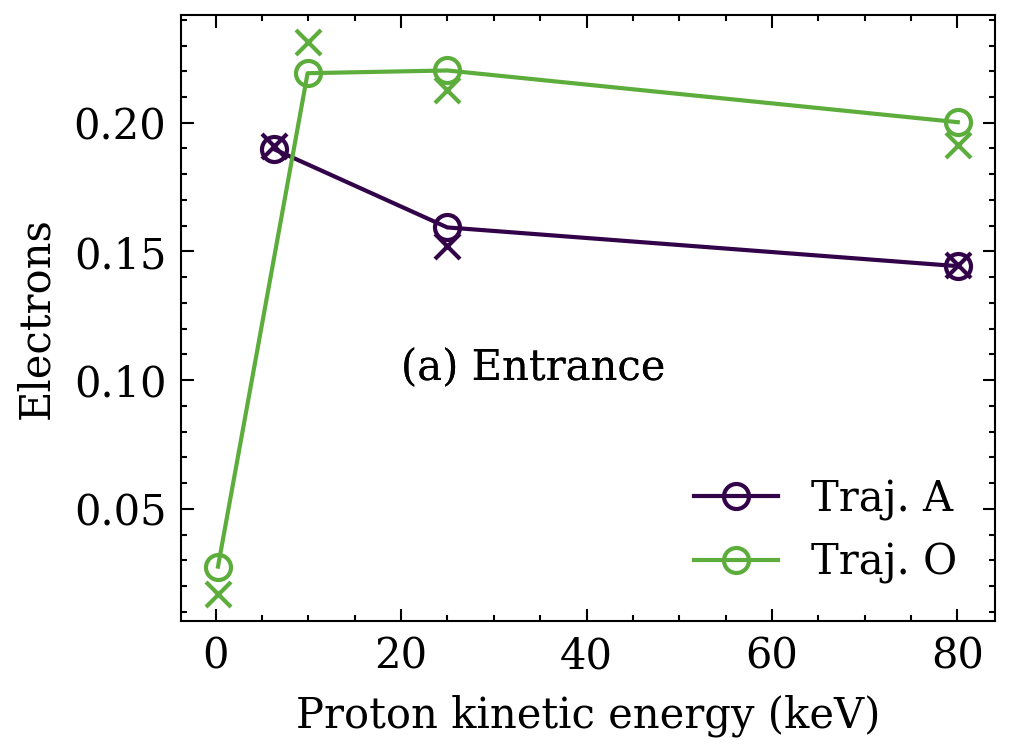}\\
\includegraphics[width=0.95\columnwidth]{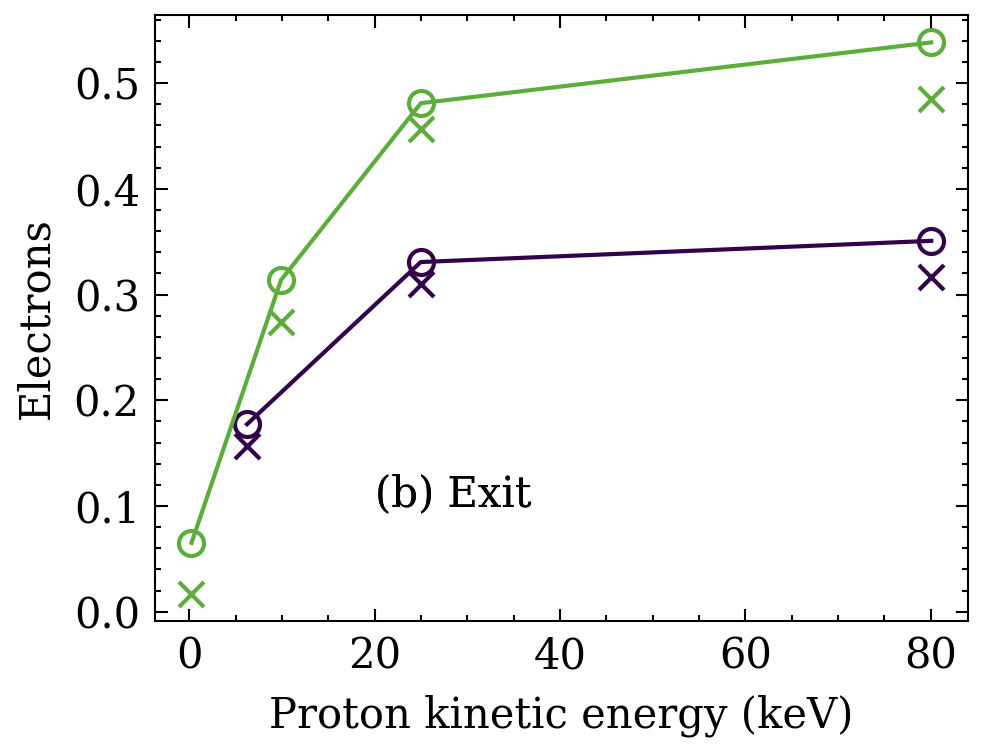}
\caption{\label{fig:seccom}
Secondary electrons from (a) the entrance side and (b) and exit side.
The solid lines with open circles are the results of this work, with absorbing boundary conditions and current density analysis.
The cross markers are the results from Ref.\ \cite{kononov_anomalous_2021}, with periodic boundary conditions and electron density analysis.
}
\end{figure}

The number of emitted electrons in the entrance- and exit-side vacuum is determined using an artificial planar detector surface, placed 10.5 $a_0$ away from the graphene.
The total number of emitted electrons $N_\mathrm{e}$ is determined by integrating the current density passing through the detector surface $\Omega$ over time,
\begin{equation}
N_\mathrm{e}=\int_\Omega j(\mathbf{r},t)\cdot \hat{\Omega} \, dt \, d\mathbf{r},
\label{eqn:total electrons}
\end{equation}
where $\hat{\Omega}$ is the normal vector of the detector plane. 
The number of electrons captured by the proton is determined by integrating the charge density over a sphere with a radius of 5 $a_0$ around the proton.
Except for computing kinetic energy spectra in Figs.\ \ref{fig:litcom} and \ref{fig:keall}, we subtract these captured electrons from the total emitted electrons in this work, because they would not be measured in experiment as secondary electrons.
Subsequently, taking the time derivative of the number of secondary electrons allows us to acquire the secondary electron current shown in Fig.\ \ref{gchar} of the main text.
These steps enable us to gain insights into the temporal behavior and pulse duration of secondary electron emission from the proton-irradiated graphene.

In Fig.\ \ref{fig:seccom}, we compare the total number of emitted electrons with prior literature data \cite{kononov_anomalous_2021} and find good agreement despite the methodological differences.
In addition, using periodic boundary conditions that allow for artificial interaction between exit-side and entrance-side electrons rather than an absorbing potential as described in Sec.\ \ref{sec:simulationcell}, Ref.\ \cite{kononov_anomalous_2021} evaluated emitted electron yields by directly integrating the electron density in the vacuum region.
Under periodic boundary conditions, charge conservation makes the electron density analysis used in Ref.\ \cite{kononov_anomalous_2021} nearly equivalent to the current density analysis of Eq.\ \eqref{eqn:total electrons}.
However, distinguishing exit-side and entrance-side electron densities requires a very long vacuum and an artificial boundary for the integration \cite{kononov_anomalous_2021,kononov_thesis_2020}.
Therefore, we consider the simulation methods used in this work (current density analysis with an absorbing potential) more accurate.

\subsection{Potential emission yield in graphene}\label{sec:pee}

We estimate the emitted electron yield due to the potential emission mechanism using the expression derived in Ref.~\cite{kishinevsky_estimation_1973},
\begin{equation}
    \gamma = \frac{0.2}{E_\mathrm{F}}\cdot \left(0.8 E_\mathrm{i} - 2 \omega_0 \right)~,
    \label{eqn:pee}
\end{equation}
where $E_\mathrm{i}$ is the ionization potential (13.6 eV for a proton) and $\omega_0$ is the work function (4.56 eV for graphene)~\cite{yu_tuning_2009, yan_determination_2012}. $E_\mathrm{F}$ is the Fermi energy of graphene, given by $E_\mathrm{F}=v_\mathrm{F}^2/2$ in atomic units, and $v_\mathrm{F}$ is measured as $1.17\cdot 10^6\, \mathrm{m/s}$~\cite{kim_direct_2012}.
Collecting all parameters, we obtained the predicted potential emission yield for proton-irradiated graphene as 0.09. 

Similarly, the ionization energy for argon is 15.76 eV~\cite{weitzel_zeke-pepico_1994}, and we ignore variations in the work function (which differs by 100 meV between monolayer and bilayer graphene~\cite{yu_tuning_2009}) and the Fermi velocity with the number of graphene layers.
The predicted potential emission yield for Ar$^+$-irradiated graphene is around 0.18.

We also note that Eq.~\eqref{eqn:pee} is most accurate in the range $3\omega_0<E_\mathrm{i}<2 (E_\mathrm{F}+\omega_0)$~\cite{kishinevsky_estimation_1973}.
The ionization potential of a proton $E_\mathrm{i}=13.6$ eV is slightly below the lower bound $3\omega_0=13.68$ eV for $\omega_0=4.56$ eV, which might result in a deviation from this model. 
Meanwhile, the argon ionization energy does fall within the model accuracy range.

\subsection{Dipole moment for plasmon excitation and plasmon decay \label{sec:SIplasmon}}

\begin{figure}[h!]
\includegraphics[width=0.95\linewidth]{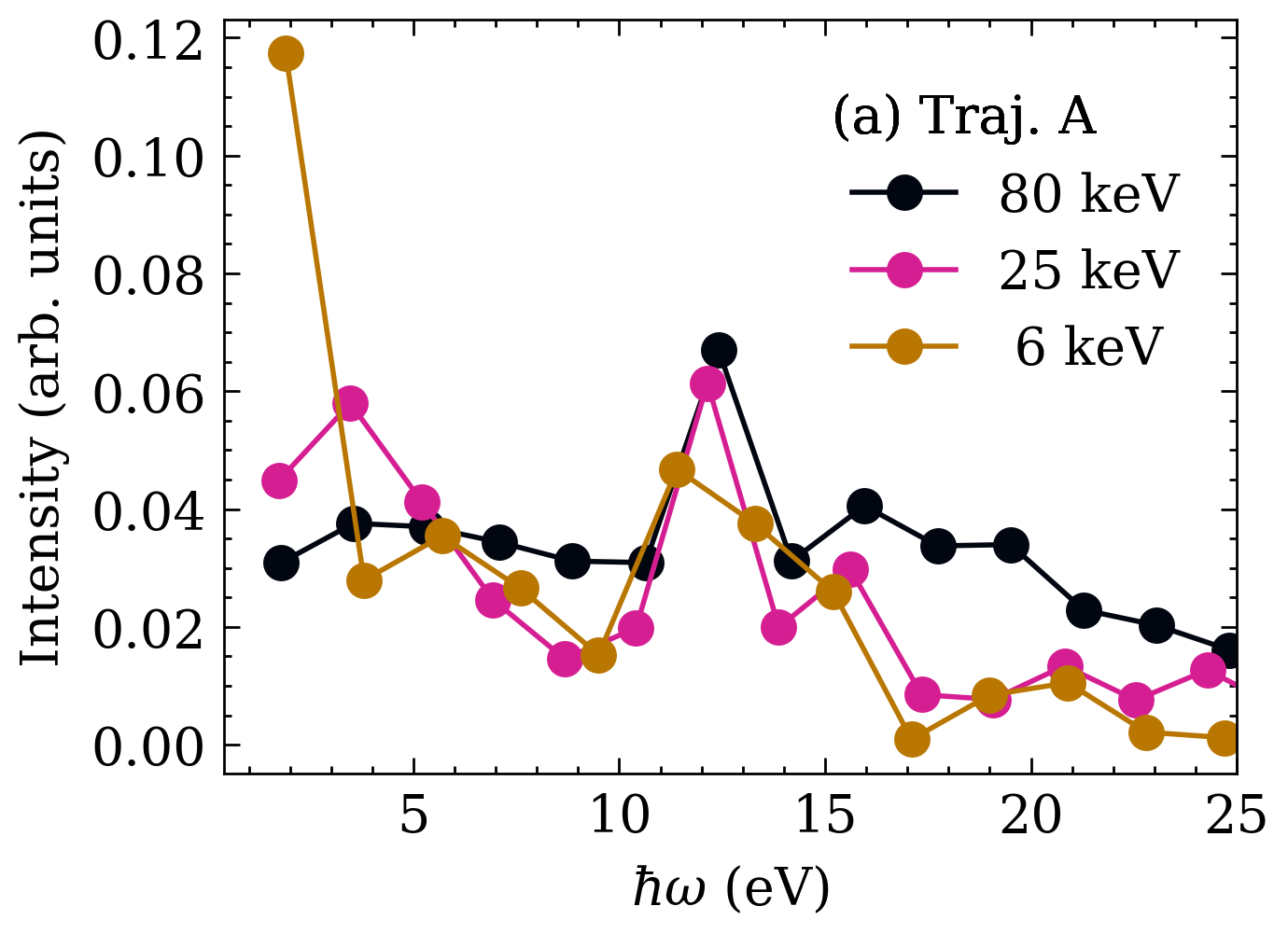} \\
\includegraphics[width=0.95\linewidth]{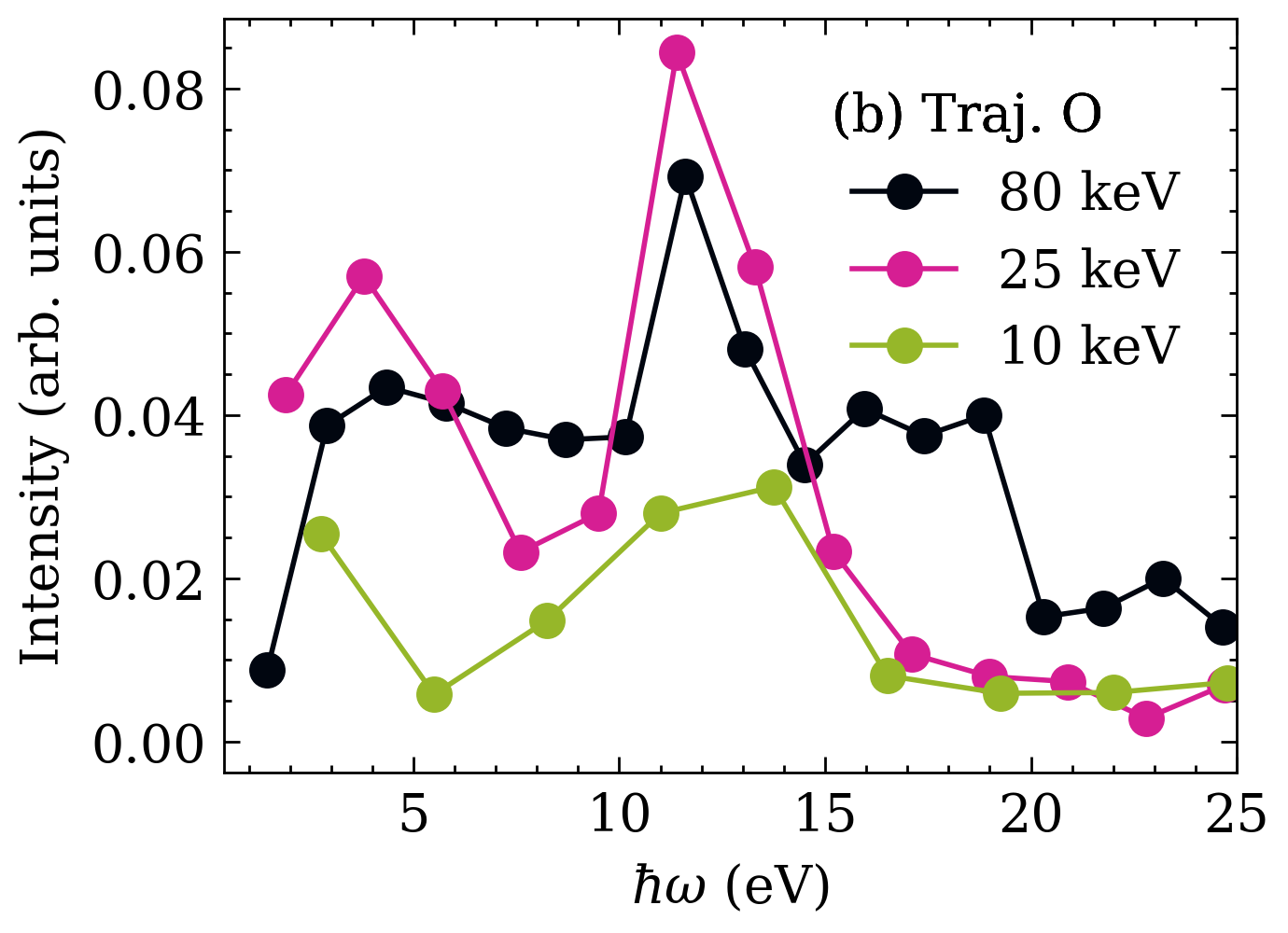}
\caption{\label{fig:dip}
Fourier transform of the dipole moment (see \mbox{Eq.\ \eqref{eqn:dipole}}) for (a) the channeling trajectory (A) and (b) the centroid trajectory (O). 
} 
\end{figure}

We calculate the dipole moment by integrating the charge density
\begin{equation}
    d(t)=\int_V z \, n(r,t) \, dr^3,
    \label{eqn:dipole}
\end{equation}
where $z$ is the position normal to the graphene surface and $V$ is the volume within $10.5 \; \mathrm{a_0}$ of graphene. To remove the influence of the electrons captured by the incident proton, we only consider times after the proton is at least 28 $\mathrm{a_0}$ away from the graphene.
To extract information about oscillation frequencies, we perform a Fourier transform of the dipole moment and obtain the data plotted in Fig.\ \ref{fig:dip}. 
Note that the coarse frequency resolution is due to the limited time-scale of the simulations.

\subsection{Kinetic energy spectra of emitted electrons from different proton trajectories and kinetic energies \label{sec:SIKEspectra}}

\begin{figure}
\includegraphics[width=0.95\columnwidth]{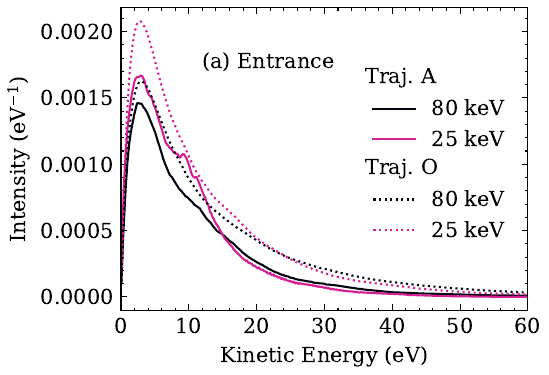}\\
\includegraphics[width=0.95\columnwidth]{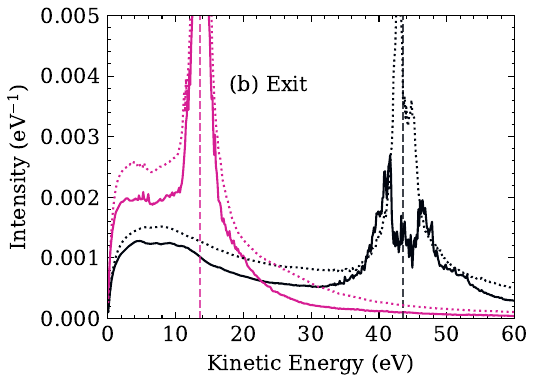}
\caption{\label{fig:keall}
The kinetic energy spectrum computed for emitted electrons, including both the secondary electrons and electrons captured by the incident proton, for (a) entrance side and (b) exit side.
Vertical dashed lines in (b) denote the electron kinetic energy corresponding to the proton velocities.
}
\end{figure}

In Fig.\ \ref{fig:keall}, we show the kinetic energy spectra of the emitted secondary electrons for several proton trajectories and kinetic energies.
In the main text, we use the kinetic energy spectra of the emitted electrons induced by the proton on the centroid trajectory and 80 keV kinetic energy as an example to compare with the experiments and Chung's model.

\begin{figure}[h!]
\includegraphics[width=0.95\columnwidth]{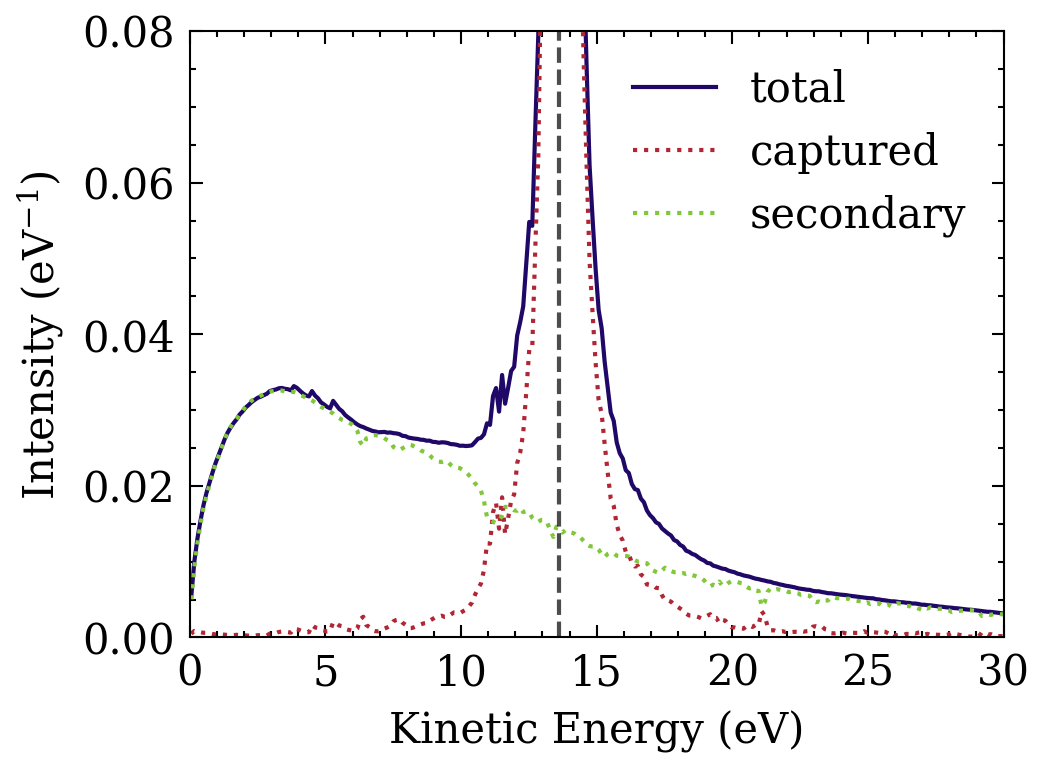}
\caption{\label{fig:KEhke}
The kinetic energy spectrum of the total emitted electrons on the exit side for the centroid trajectory (O) and 25 keV proton kinetic energy, following Eq.~\eqref{eq:simpleint}, with time after impact 1.5 fs. The secondary electrons are the total emitted electrons minus the captured electrons. The captured electrons, which are the electrons within a radius of 5 $a_0$ around the proton, form a peak at the proton velocity (vertical dashed line).  
}
\end{figure}

The kinetic energy spectrum we compute for the exit side (see Fig.\ \ref{fig:keall}b) shows a clear peak that corresponds to electrons with a kinetic energy that matches the proton velocity.
Those electrons are mostly located around the proton and we compare the kinetic energy spectrum of electrons within 5 $a_0$ of the proton to the entire spectrum in Fig.\ \ref{fig:KEhke}.
From this, we conclude that these electrons are captured by the proton and would not be detected in the experimentally measured secondary electron spectrum.
While it is possible that captured electrons could be detached if they are weakly bound, this phenomenon is not observed in our few-fs RT-TDDFT simulations.
Also, these captured electrons mainly occupy the 1$s$ orbital of the proton~\cite{kononov_anomalous_2021,kononov_anomalous_2021_data}, indicating that most of them are strongly bound.
In contrast, highly charged ions can lose captured electrons due to autoionization~\cite{wilhelm_interatomic_2017, creutzburg_vanishing_2020}. 

Second, the kinetic energy spectra in Fig.\ \ref{fig:keall} also show a characteristic peak of around 3 eV for the entrance side and around 5 eV for the exit side.
The overall energy position of this peak and the general shape of the spectra in this kinetic energy region are very similar among all trajectories and proton velocities on both the entrance and exit sides.
Varying incident proton energy and trajectory primarily influences the amount of emitted electrons, but only slightly alters their kinetic energy distribution. 
This implies that the emission process is predominantly governed by intrinsic material properties of the graphene target, such as the mean free path of excited electrons and the work function.

Moreover, Fig.\ \ref{fig:keall} provides a more detailed understanding of the secondary electron pulse observed in the real-time evolution shown in Fig.\ \ref{gchar}.
From the real-time data of emission current at the detector surface (see Fig.\ \ref{gchar}) it is hard to disentangle which of the emitted electrons travel faster than others and when they were emitted.
The kinetic energy spectrum shows that there is no clear shift of the peak position between the centroid trajectory (O) and the channeling trajectory (A). 
Although the centroid trajectory (O) has a larger magnitude of the high-energy tail, it might not be large enough to explain the earlier onset of the secondary electron current in Fig~\ref{gchar}. 

The short time scale of secondary electron emission shown in Fig.\ \ref{gchar} indicates that delayed electron emission, resulting from the relaxation of the energy stored during ion impact, is not observed in our simulations.
In particular, while the dipole moment of the graphene after proton impact shows distinct plasmonic oscillations (see Fig.~\ref{fig:dip}), plasmon decay is not seen in our real-time profile of the secondary electron emission.
Plasmon decay with an intrinsic plasmon lifetime~\cite{ni_fundamental_2018} of about 12 ps is much slower than the timescale of our simulations.
If the time-scale in our RT-TDDFT simulation were long enough to capture such delayed processes
we might be able to observe a shoulder-like feature as experimentally detected in the kinetic energy spectrum of the secondary electrons emitted from ion-irradiated metallic surfaces \cite{baragiola_plasmon_2001, baragiola_plasmon-assisted_1996}. 
Plasmon-assisted emission was also reported as a local minimum in the derivative of the kinetic energy spectrum near $\omega_p-\omega_0$, where $\omega_p$ is the plasmon frequency and $\omega_0$ is the work function of host materials~\cite{baragiola_plasmon_2001, baragiola_plasmon-assisted_1996, riccardi_bulk_2003, riccardi_effects_2022}. 
In the future, we envision comparing with experiments in more detail
to quantify the electron yield due to delayed processes such as plasmon decay. 

We note that the two emission mechanisms discussed above, i.e.\ potential and kinetic emission, contribute differently for different projectiles.
For example, Xe$^{40+}$ is a highly charged ion and is expected to be dominated by potential emission~\cite{schwestka_charge-exchange-driven_2019}, while potential emission in our proton simulations is less important, as discussed above.
The good agreement of the kinetic energy spectra for these very different projectiles indicates that electron emission is an intrinsic property of the target that is not influenced by the different types of projectiles.

Several reasons might lead to our simulated kinetic energy spectrum differing from the experimental spectrum.  
One possible error is that our approach may not fully account for long-term secondary electron dynamics near the surface due to the finite-size effects of our finite vacuum size, and hence the peak position might be overestimated (see Fig.~\ref{fig:KEvLtimeall} for the convergence of spectra over time). 
Besides, due to the finite size of the vacuum in our simulation, the emitted electrons might experience complicated forces from both the graphene and the proton. 
Hence the emitted electrons might experience some acceleration from the fast proton or deceleration from the positively charged graphene.

\subsection{Convergence of kinetic energy spectrum}\label{sec:ke spectrum}

\begin{figure}[h!]
\includegraphics[width=0.95\columnwidth]{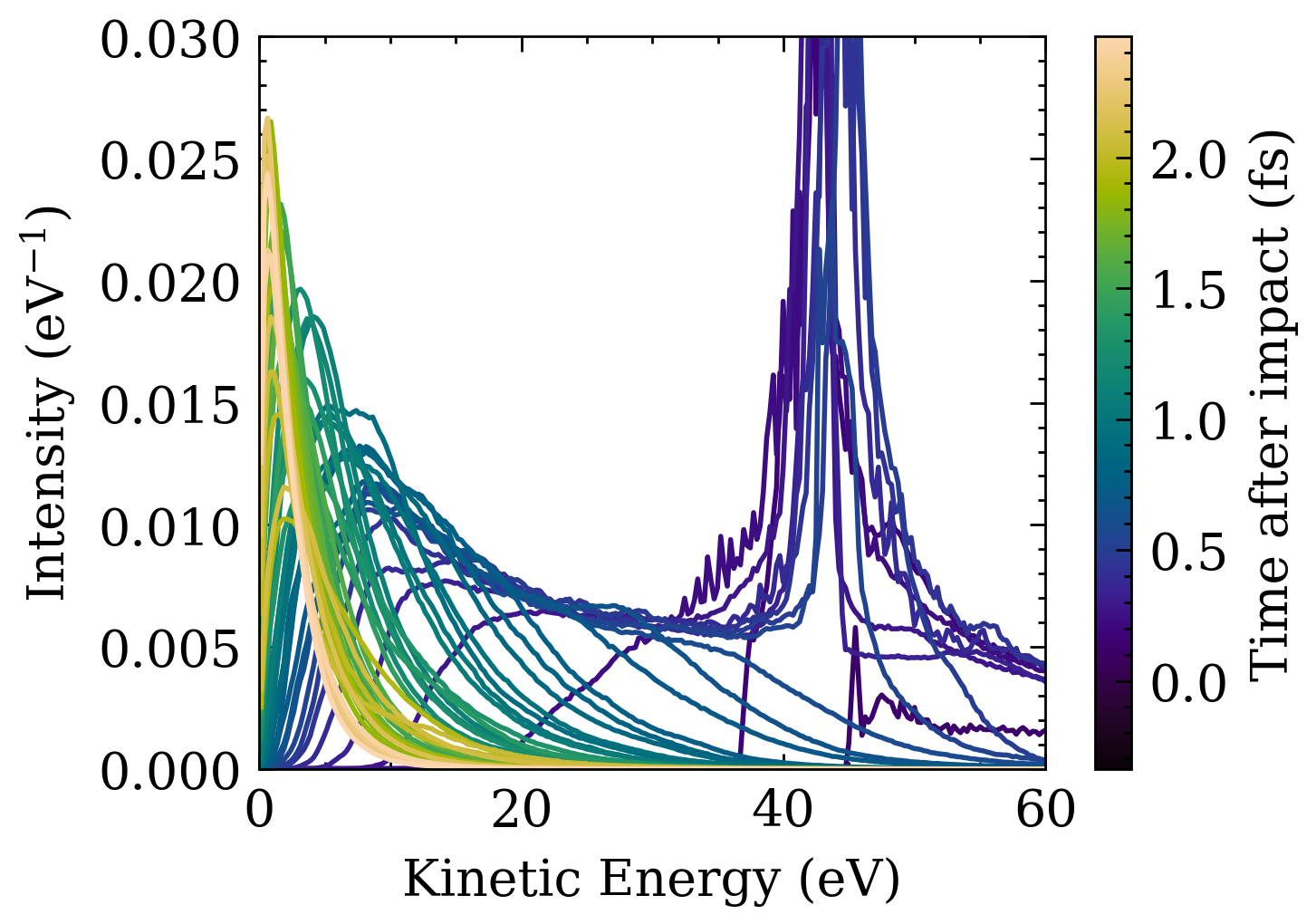}
\caption{\label{fig:KEtime}
The time-dependent kinetic energy spectra of the exit side of the centroid trajectory (O) and 80 keV proton kinetic energy, calculated by Eq.~\eqref{eq:ke hist}, are shown in solid lines.
}
\end{figure}

\begin{figure}[h!]
\includegraphics[width=0.95\columnwidth]{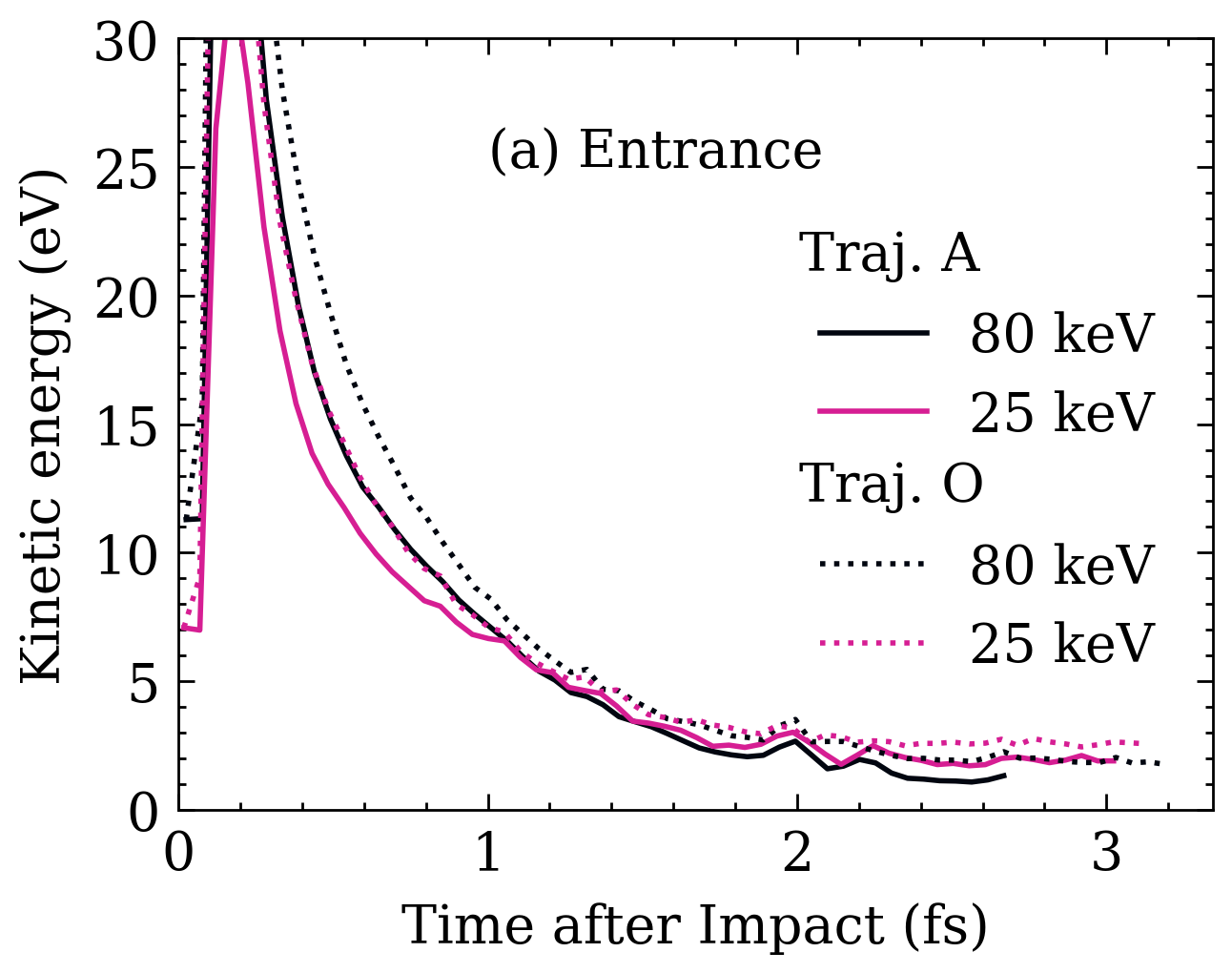}\\
\includegraphics[width=0.95\columnwidth]{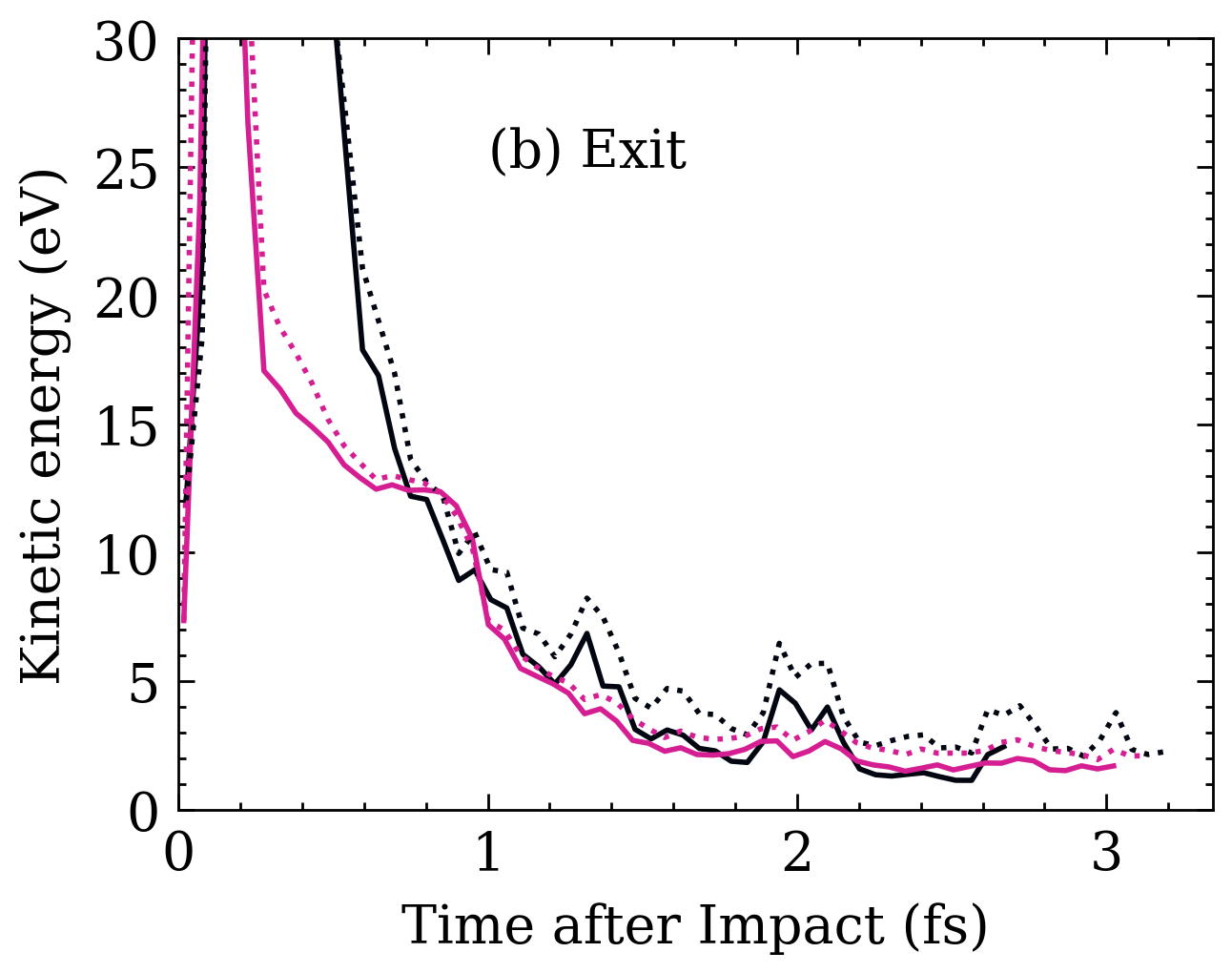}
\caption{\label{fig:KEavetime}
The average of the instantaneous kinetic energy spectrum of the electrons in the vacuum in (a) the entrance side and (b) the exit side. After around 1.5 fs, the average kinetic energy of the electrons in the vacuum remains around 2 eV.
}
\end{figure}
  
\begin{figure}
\includegraphics[width=0.95\columnwidth]{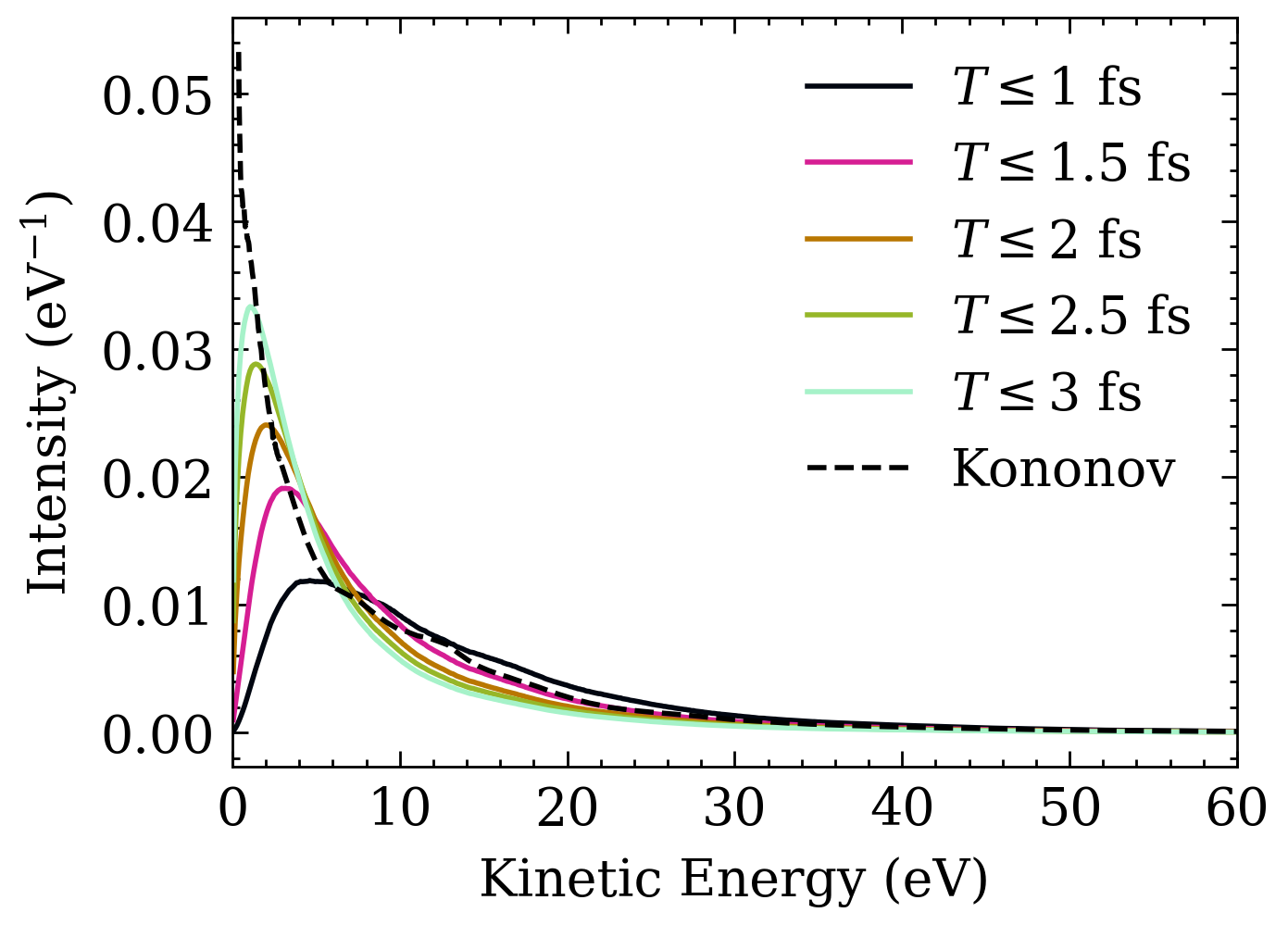}\\
\includegraphics[width=0.95\columnwidth]{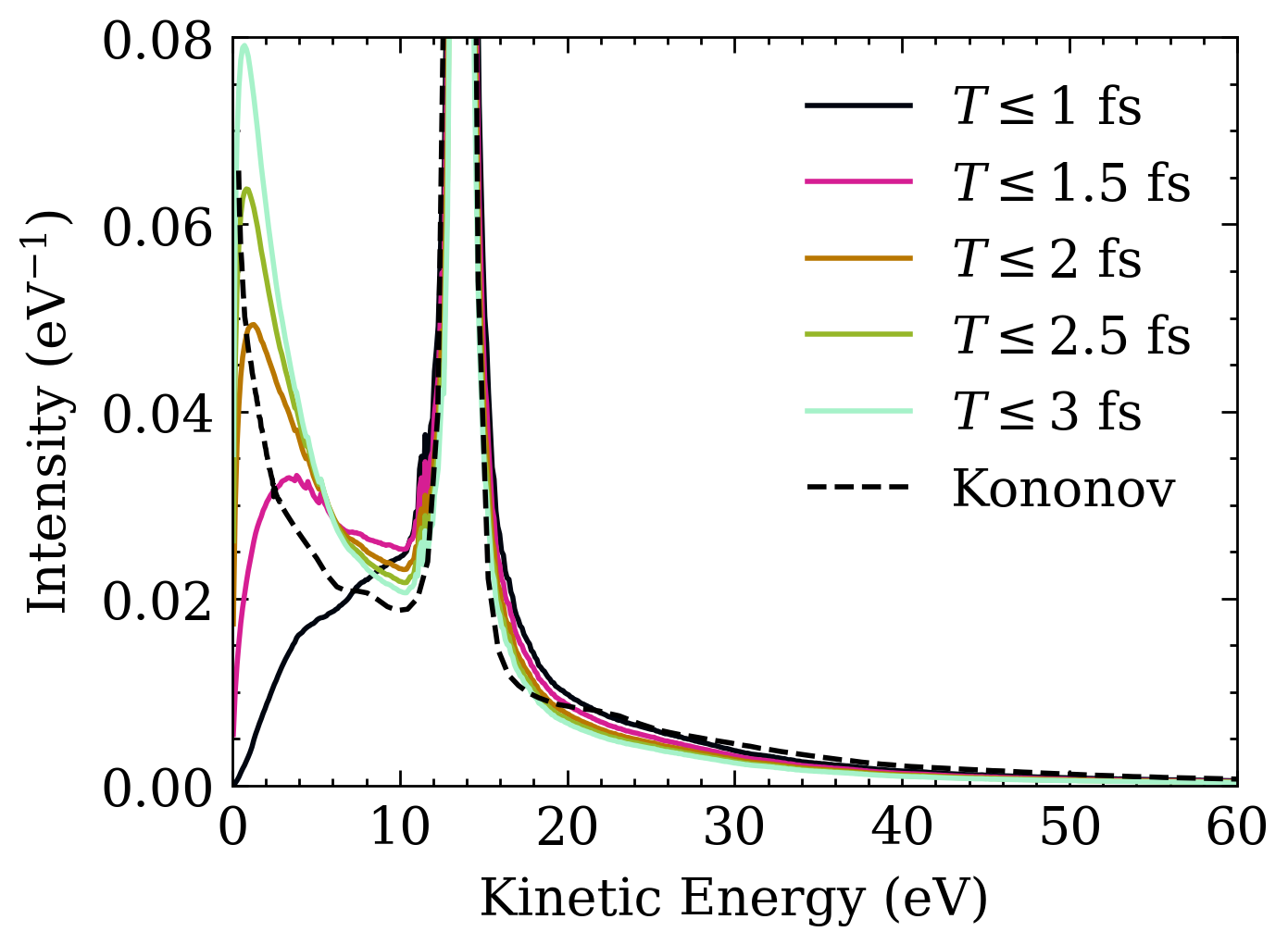}
    \caption{\label{fig:KEtimeall}
    Comparison of the kinetic energy spectrum of centroid trajectory (O) and proton kinetic energy (25 keV) with the different cutoff time with the kinetic energy spectrum calculated by the time-of-flight analysis~\cite{vazquez_electron_2021,kononov_thesis_2020} in entrance side (top), and exit side (bottom).
    }
\end{figure}

One problem with our kinetic energy spectrum calculation is that a single instantaneous spectrum is not sufficient to capture the kinetic energy distribution of all the emitted electrons (see Fig.~\ref{fig:KEtime}).
The average kinetic energy of the emitted electrons remaining in the vacuum begins decaying shortly after impact (see Fig.~\ref{fig:KEavetime}) because the absorbing potential removes faster electrons over time while slower electrons require more time to emerge into the vacuum.

We solve this problem by accumulating the instantaneous kinetic energy spectrum of the electrons in the vacuum over time. 
To motivate our approach, we first consider simply integrating the instantaneous spectrum as
\begin{equation}
P_\mathrm{acc}(\mathrm{KE}) = N_\mathrm{KE} \int_0^T P(\mathrm{KE},t) \, dt,
\label{eq:simpleint}
\end{equation}
where $N_\mathrm{KE}$ is a normalization factor that ensures that $P_\mathrm{acc}(\mathrm{KE})$ describes the total number of electrons emitted within time after impact $T$, calculated by Eq.~\eqref{eqn:total electrons}.
Eq.~\eqref{eq:simpleint} would count emitted electrons proportionally to how much time they spend within the vacuum, exaggerating low-energy features of the spectrum and underweighting high-energy portions.
For this reason, Eq.\ \eqref{eq:simpleint} also converges slowly with increasing $T$ (see Fig.\ \ref{fig:KEtimeall}), as slow electrons with kinetic energy on the order of 1 eV remain in the vacuum for at least 3 fs.

\begin{figure}
\includegraphics[width=0.95\columnwidth]{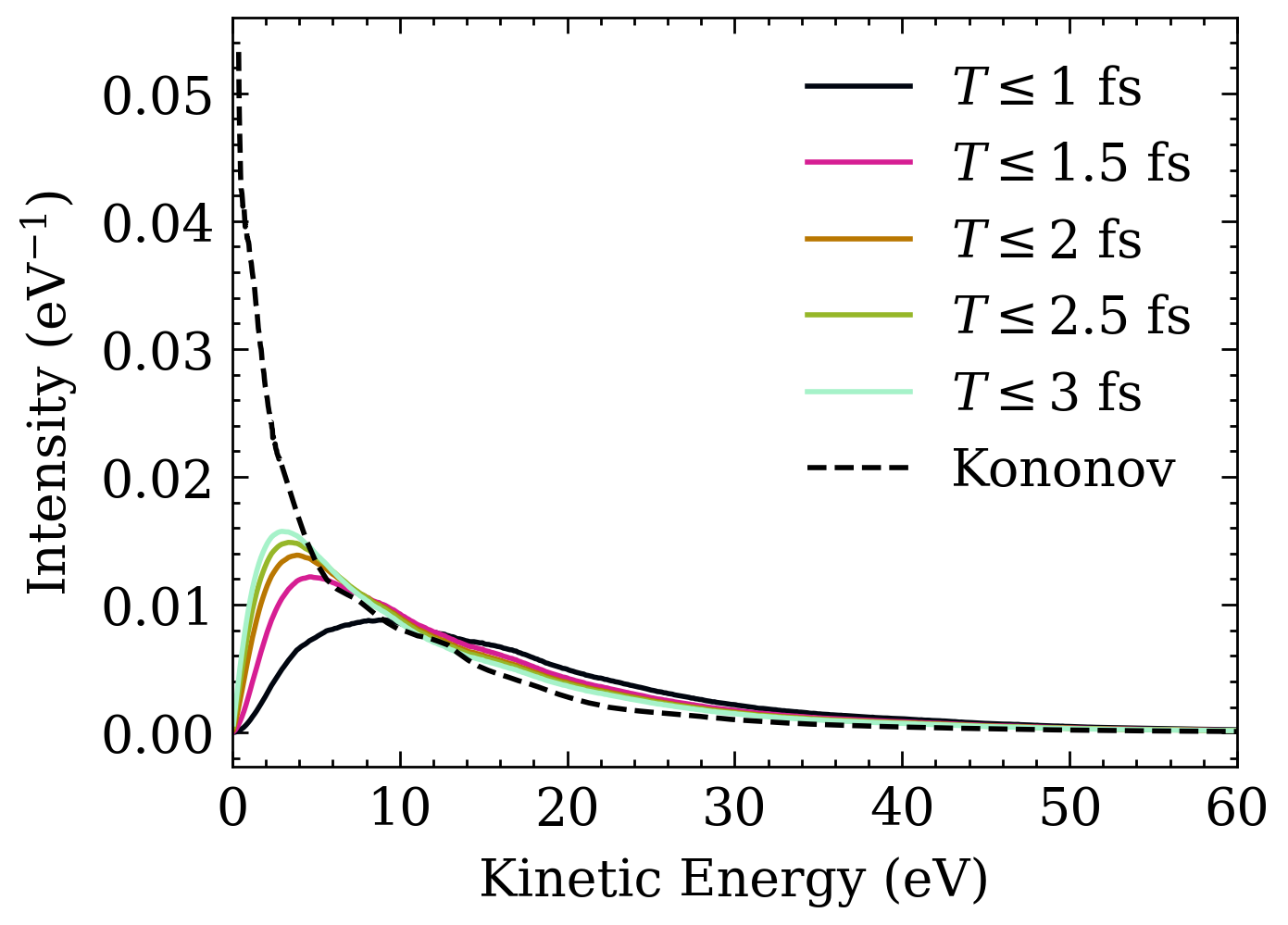}\\
\includegraphics[width=0.95\columnwidth]{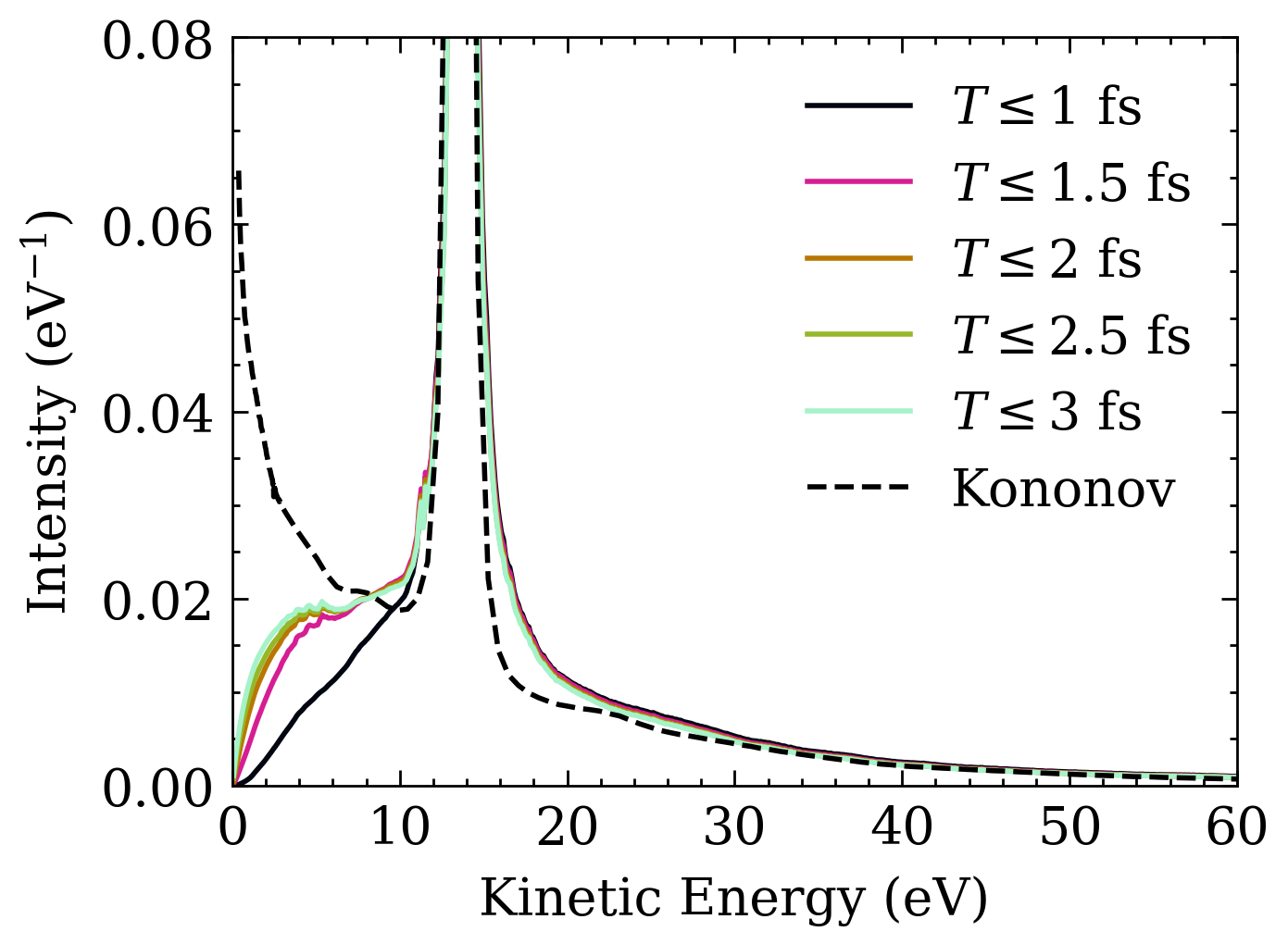}
    \caption{\label{fig:KEvLtimeall}
    Comparison of the kinetic energy spectrum of centroid trajectory (O) and proton kinetic energy (25 keV) with the different cutoff time, weighted by $v_z/L$, with the kinetic energy spectrum calculated by the time-of-flight analysis~\cite{vazquez_electron_2021,kononov_thesis_2020} in entrance side (top), and exit side (bottom).
    }
\end{figure}

To correct Eq.~\eqref{eq:simpleint} for the varying amount of time that electrons with different speeds spend within the simulation, we modify Eq.~\eqref{eq:ke hist} with the weighting function $w(\mathbf{r},t)$
\begin{equation}
P(\mathrm{KE},t)= \int_{V_{\mathrm{vac}}} w(\mathbf{r},t) \, n(\mathbf{r}, t) \, \delta\left(\mathrm{KE}-\frac{v^2(\mathbf{r}, t)}{2}\right) \, d\mathbf{r},
\label{eq:correctedint}
\end{equation}
where 
 \begin{equation}
    w(\mathbf{r},t) = \frac{v_z(\mathbf{r},t)}{L}.
\end{equation}
    
Here $v_z$ is the $z$ component of the velocity of the emitted electrons, and $L=40\, a_0$ is the distance from the graphene surface to the absorbing potential.
The resulting kinetic energy spectrum converges after about 1.5 fs (see Fig.~\ref{fig:KEvLtimeall}), consistent with the length of the secondary electron pulse shown in Fig.~\ref{gchar} of the main text.

\begin{figure}
    \centering
    \includegraphics{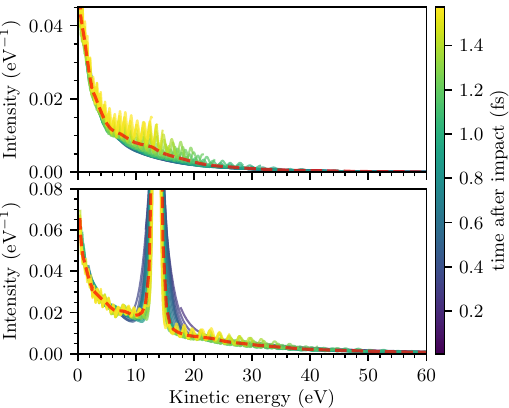}
    \caption{
    Instantaneous entrance-side (top) and exit-side (bottom) kinetic energy spectra obtained from time-of-flight analysis used in Refs.~\onlinecite{vazquez_electron_2021,kononov_thesis_2020} for a 25 keV proton with the centroid trajectory.
    The red dashed curves indicate time-averaged data according to Eq.~\eqref{eq:tofint}.
    }
    \label{fig:KEtof}
\end{figure}

The TOF analysis \cite{ullrich_time-dependent_2012} applied in earlier work \cite{vazquez_electron_2021,kononov_thesis_2020} suffers from similar challenges with appropriately combining high-energy data accessed shortly after impact and low-energy data only accessed at later times.
Rather than using the current density, TOF analysis infers an electron velocity distribution using only the electron density by assuming a particular spatiotemporal origin for electron emission, in this case, the graphene plane at the time of proton impact.
In principle, this assumption becomes more accurate over time, as any deviations from the assumed spatiotemporal origin become smaller relative to the elapsed time and the distance traveled by emitted electrons.
To preferentially weigh more converged data at later times while retaining information about high-energy electrons at early times, Refs.~\onlinecite{vazquez_electron_2021,kononov_thesis_2020} modified Eq.~\eqref{eq:simpleint} with a time-dependent weighting function $w'(t)$,
\begin{equation}
P_\mathrm{acc}(\mathrm{KE}) = \frac{1}{\int_0^T w'(t) \, dt} \int_0^T w(t) \, P(\mathrm{KE},t) \, dt,
\label{eq:tofint}
\end{equation}
and arbitrarily chose $w'(t) = t$ to obtain the time-averaged spectra shown in Fig.~\ref{fig:KEtof}.
A velocity-dependent weighting function as in Eq.~\eqref{eq:correctedint} was not needed because the earlier studies did not include an absorbing potential and thus emitted electrons remained within the vacuum indefinitely.

Figure \ref{fig:KEvLtimeall} compares our new method for computing kinetic energy spectra to the prior TOF results~\cite{vazquez_electron_2021,kononov_thesis_2020}.
Although the spectra show some agreement --- e.g., at energies $\gtrsim$ 5 eV in the entrance-side spectrum and near the exit-side feature corresponding to captured electrons --- we note several important qualitative differences.
First, the TOF data underestimates the high-energy tail because high-energy electrons artificially slow down over time as they approach the periodic boundary with no absorbing potential.
In addition, the TOF analysis assumes electrons are only emitted perpendicularly to the graphene.
By neglecting in-plane velocity components that become increasingly significant for slow electrons, the TOF data further overestimates the low-energy region of the spectrum.
Overall, we expect that our new technique for computing kinetic energy spectra using the current density and an absorbing potential is more accurate, and we use this method for the results presented in the main text.

\subsection{Real-time electron emission of different trajectories at different lattice temperatures}

\begin{figure}[h!]
\includegraphics[width=0.95\columnwidth]{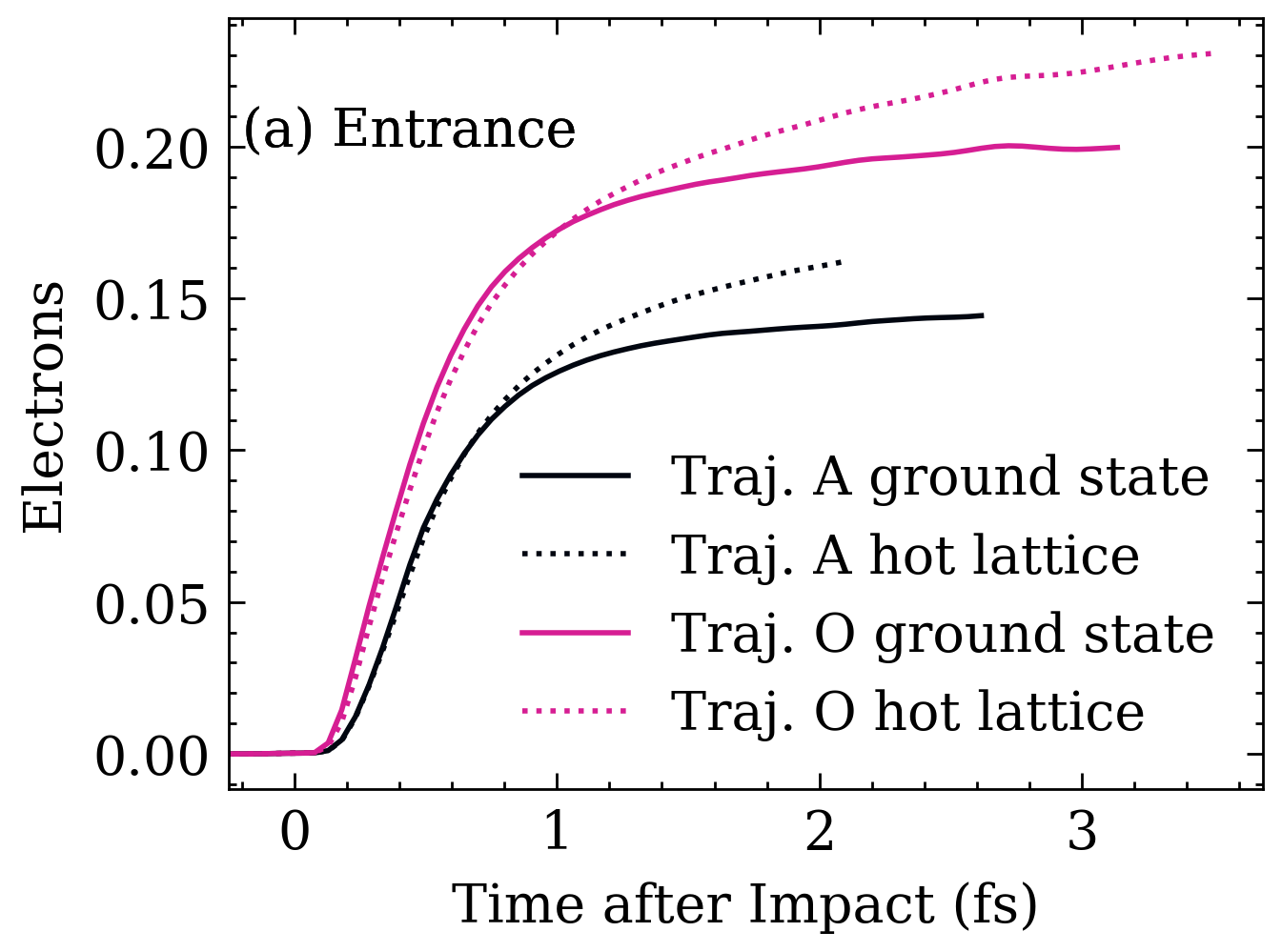}\\
\includegraphics[width=0.95\columnwidth]{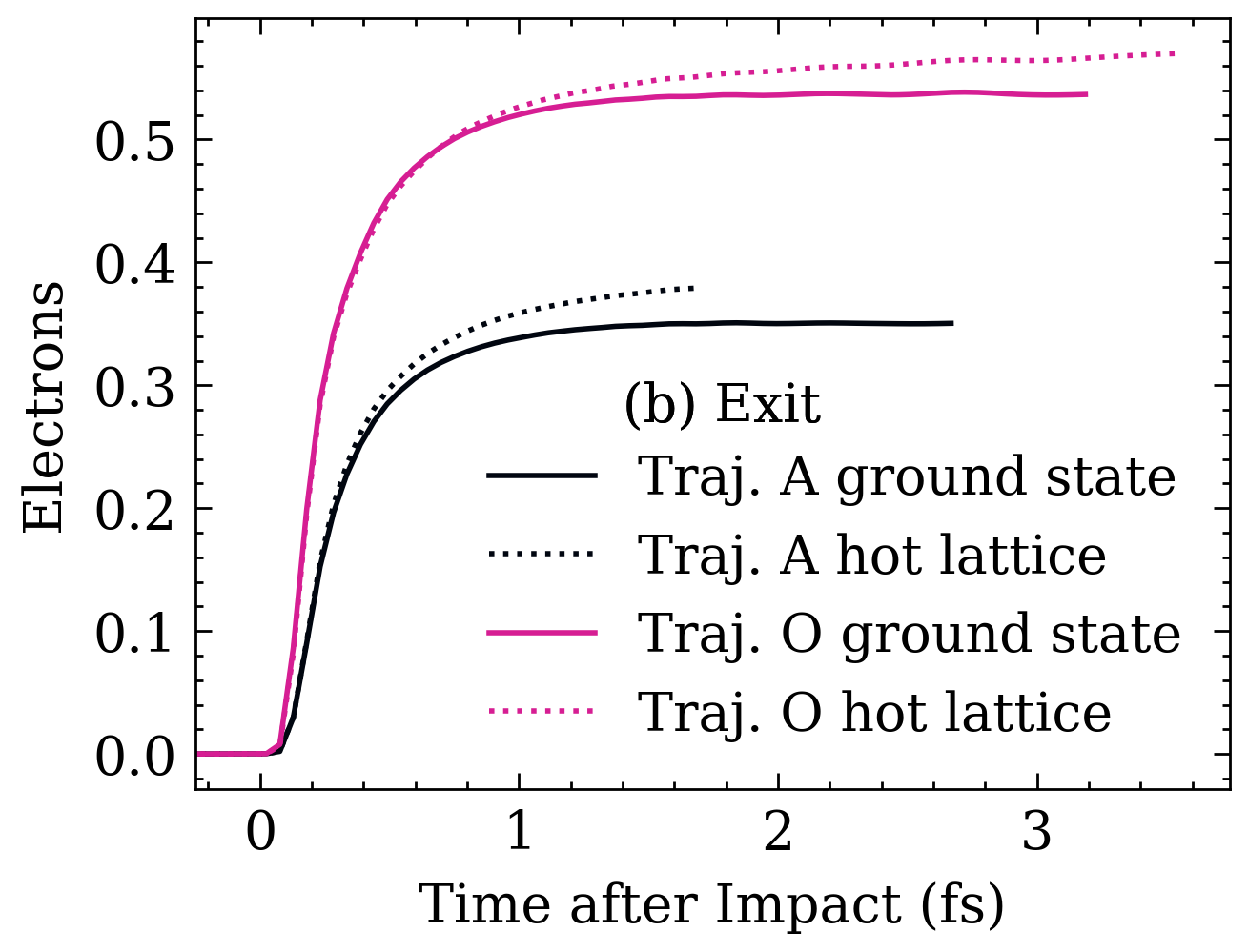}
\caption{
\label{fig:TAOcom}
Electron emission at the elevated lattice temperature of $1,000$ K, compared to emission at ground state 0 K, in both channeling trajectory (A) and centroid trajectory (O).
The proton is exactly in the graphene plane at ``impact''.}
\end{figure}

In the main text, we demonstrate the enhancement of secondary electron emission at the elevated lattice temperature under the centroid trajectory (O). 
To further examine the trajectory dependence of such enhancement of electron emission, we investigate the electron emission at the elevated lattice temperature under the channeling trajectory (A). 
Figure \ref{fig:TAOcom} demonstrates that an enhanced emission also occurs in the channeling trajectory (A) at this elevated lattice temperature,  similar to the centroid trajectory.

\subsection{Kinetic energy spectrum at elevated lattice temperature}

\begin{figure}[h!]
\includegraphics[width=0.95\linewidth]{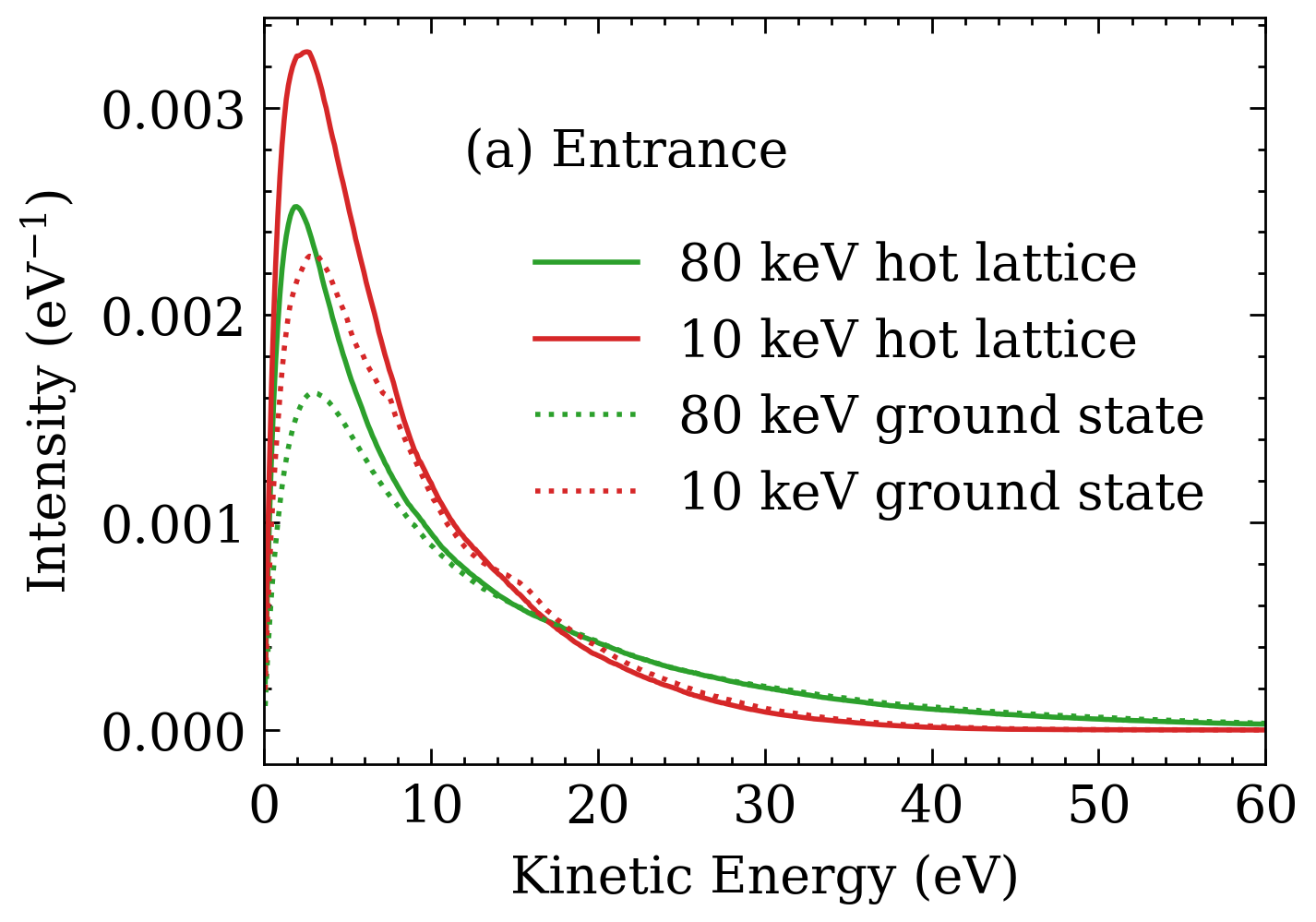} \\
\includegraphics[width=0.95\linewidth]{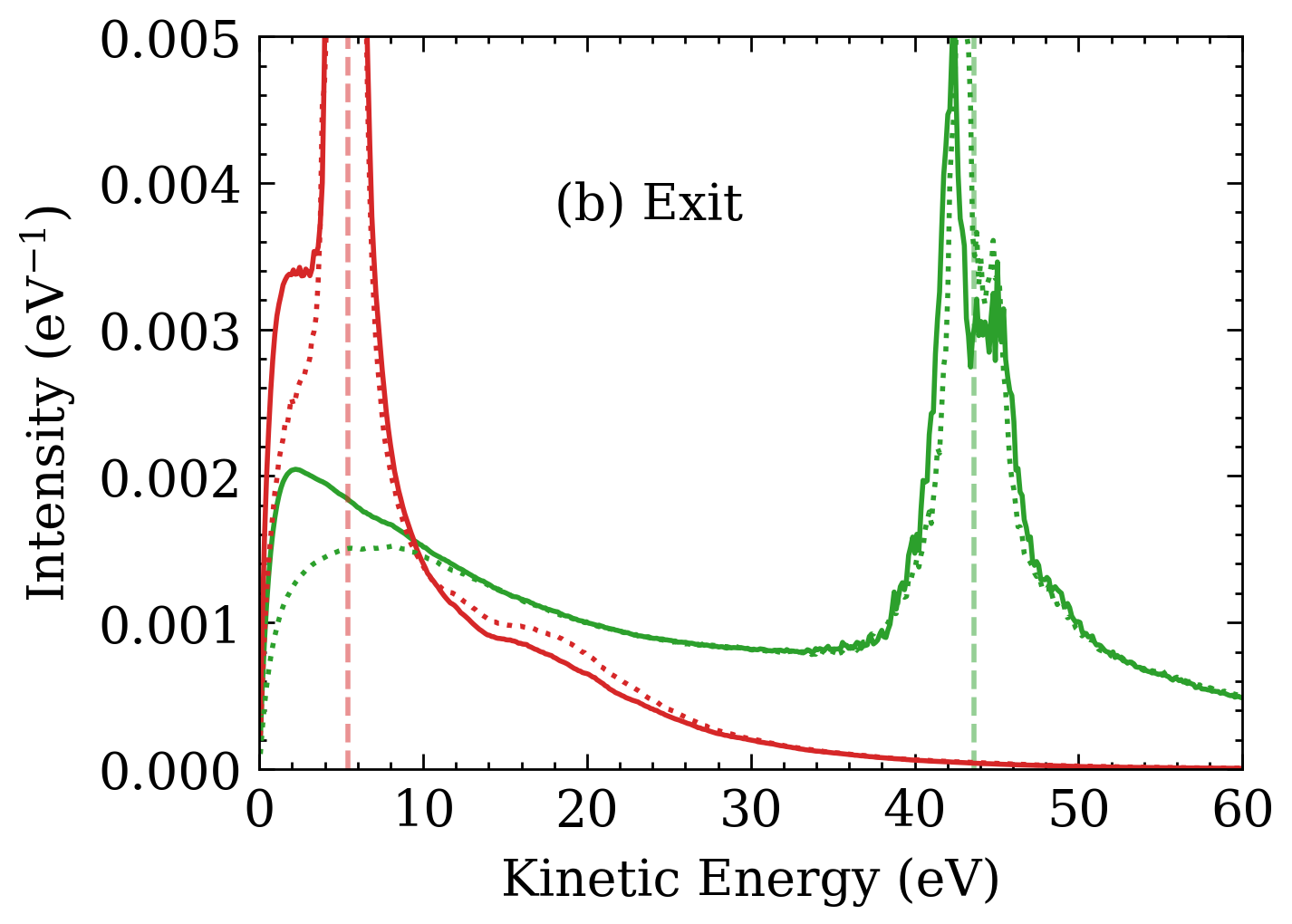}
\caption{\label{fig:ClatticeKE}
The kinetic energy spectrum of the emitted electrons of the centroid trajectory (O) at lattice temperature 1000 K, compared with the kinetic energy spectrum at the ground state 0 K lattice temperature. The vertical dashed lines indicate the kinetic energy of the electrons moving at the same speed as the proton. 
} 
\end{figure}

In the main text, we demonstrate that the elevated lattice temperature results in an enhanced secondary electron emission. 
Figure \ref{fig:ClatticeKE} demonstrates that secondary electron emission from graphene at the elevated lattice temperature still has a similar kinetic energy distribution, compared to the emitted electrons from a graphene at zero lattice temperature. 
Hence, we suggest using secondary electron yield, but not the kinetic energy distribution, as a probe of the electron dynamics in the materials.

\end{document}